\documentclass[aps, prl, twocolumn, reprint, superscriptaddress]{revtex4-2}
\usepackage{graphicx} 
\usepackage{booktabs}
\usepackage{subcaption}
\usepackage{float}
\usepackage{amsmath}
\usepackage{multirow}
\usepackage{textgreek}
\usepackage{dcolumn}
\usepackage{placeins}
\usepackage{bm}

\usepackage{array}
\newcommand{\PreserveBackslash}[1]{\let\temp=\\#1\let\\=\temp}
\newcolumntype{C}[1]{>{\PreserveBackslash\centering}p{#1}}
\newcolumntype{R}[1]{>{\PreserveBackslash\raggedleft}p{#1}}

\begin{document}

\title{Accurate and Efficient Interatomic Potentials for Dislocations in InP}
\author{Thomas Rocke}
\affiliation{Warwick Centre for Predictive Modelling, School of Engineering, University of Warwick, Coventry, CV4\,\,7AL, UK}
\author{Thomas Hudson}
\affiliation{Warwick Mathematics Institute, Zeeman Building, University of Warwick, Coventry, CV4\,\,7AL, UK}
\author{Richard Beanland}
\affiliation{Department of Physics, University of Warwick, Coventry, CV4\,\,7AL, UK}
\author{James Kermode}
\affiliation{Warwick Centre for Predictive Modelling, School of Engineering, University of Warwick, Coventry, CV4\,\,7AL, UK}

\date{April 2026}

\begin{abstract}
We present Atomic Cluster Expansion (ACE) and MACE models trained on a new dataset of Density Functional Theory (DFT) calculations, constructed for the task of studying the mobility of dislocations in Indium Phosphide (InP). The models are validated in a suite of tests against RSCAN DFT, and compared with previously published potentials from literature. Our new models act as much better surrogates for DFT than the literature models: errors on partial dislocation formation energies are at most 4\% for both ACE and MACE, compared with 18\% for the MACE-MPA foundation model and 42–50\% for earlier bespoke potentials. The bespoke MACE model achieves this accuracy while being around five times faster to evaluate than the MP0 and MPA foundation models.
\end{abstract}

\maketitle

\section{Introduction}
Indium Phosphide (InP) is a stoichiometrically simple example of a III-V semiconductor, a family of materials which are commonly used in optoelectronic devices. Dislocations play a key role in the failure and degradation mechanisms for such devices \cite{Device1, Device2}, and so having accurate tools which are scalable to large dislocation systems is vital to understanding device failure, as well as enabling future improvements to devices. 

Early attempts to model dislocation properties using Density Functional Theory (DFT) have used very small dipolar \cite{DislocDipole}, quadrupolar \cite{DislocQuad1, DislocQuad2} or hydrogen-terminated cylindrical \cite{DislocCyl} arrangements. Later studies used flexible boundary conditions \cite{DislocGreens1, DislocGreens2}, or coupled Quantum Mechanics/Molecular Mechanics (QM/MM) simulations \cite{QMMM1, QMMM2, QMMM3} (i.e. coupling a region governed by DFT to one governed by an interatomic potential) in order to reduce the finite size effect of the QM system by converging the energy stored in the dislocation strain field. By training a machine learning interatomic potential (MLIP) model to be a surrogate model for DFT, we could model dislocations in InP directly in a single large cylinder without needing complex boundary conditions or coupling schemes.

We compare our new models against the Branicio et al Vashishta potential \cite{Vashishta} and the SNAP potential \cite{SNAP} by Cusentino et al, which are both bespoke models for InP. The Vashishta model was parameterised to reproduce experimental observations of zincblende InP, including the lattice constant, cohesive energy, melting temperature, bulk modulus, and the C$_{11}$ and C$_{12}$ elastic constants. The SNAP model was fitted to a dataset of LDA DFT calculations, covering bulk elasticity, the equation of state, and point defect formation.

We also test the medium ``MP0" and ``MPA" MACE Foundation models \cite{MP0} which attempt to be universally applicable to any material system. The MP0 model was trained on the Materials Project dataset \cite{MaterialsProject}, and the MPA model used the same dataset plus the addition of the sAlex dataset \cite{sAlex}, which is a subsampled version of the Alexandria dataset \cite{Alexandria}.

The methodology section describes the process of constructing the dataset and training the models, including the DFT parameterisation, contents of the dataset, and the hyperparameterisation of the MLIP models. The results section then describes the set of benchmarks used to compare the new models with literature alternatives, and shows the results of these benchmarks. Models validated with this suite of tests would then be suitable for future work exploring large-scale modeling of dislocations in indium phosphide.

\section{Methodology}
\label{sec:Method}
\subsection{Dataset Design}
In order to generate stable and accurate models for studying dislocations, the dataset needs to provide representative atomic environments which strongly correlate to atomic environments present in the dislocation structures we wish to apply the models to. 

Firstly, the dataset needs good coverage of the zincblende bulk. Including strained structures in the linear elastic regime is an important first step in accurately describing the long-range strain field of a dislocation. 

Secondly, coverage over point defects is needed, as the interaction between point defects and the dislocation core is known to be important to the dislocation climb mechanism \cite{DislocPD1, DislocPD2, DislocPD3}. The defects covered include antisites, vacancies, and the most thermodynamically stable interstitials (tetrahedral and octahedral sites for In, and the dumbbell site for P), with a particular bias towards vacancies and interstitials. The dataset includes structures targeting both the formation and the migration of these defects.

Thirdly, the dataset provides coverage over the (001), (011) and (111) stacking faults. Including the (111) stacking fault is vital to modeling dissociated dislocations, where two partial dislocation cores are connected by this stacking fault. The other stacking faults provide varied strain fields which can improve model stability, although they are less common in InP crystals.

Finally, dislocations are included in the form of dislocation quadrupole structures. These are small, periodic structures which contain two dislocation cores of opposite Burgers vector, which allow dislocations to be studied with DFT directly, without the need for free or hydrogenated surfaces of previous works. Included are quadrupoles of the $60^\circ$ and screw dislocation cores, and the related $30^\circ$ and $90^\circ$ partial dislocation cores \cite{DiamondPartials, InPPartials}.

The dataset aims to produce robust potentials by drawing samples taken from  Molecular Dynamics (MD), as well as random perturbations of the atomic positions and cell geometry, starting from structures minimised by previous iterations of potentials. The MD all used Langevin dynamics, with temperatures up to 500 K (mainly 300 K). Random cell perturbations were generated by sampling from the distribution
\begin{equation}
    \begin{pmatrix}
        \epsilon_{xx}, \epsilon_{yy}, \epsilon_{zz}, \gamma_{yz}, \gamma_{xz}, \gamma_{xy}
    \end{pmatrix} \sim \mathcal{N}(0, 0.01^2)
\end{equation}
and atomic positions were displaced using samples from a normal distribution of standard deviation 0.1\,Å.

The dataset makes use of the Non-Diagonal Supercell (NDSC) method \cite{NDSC1, NDSC2}, which was initially applied to MLIP development in order to efficiently facilitate the learning of elastic and phononic properties. Here, we also apply the technique to cells containing point defects in order to attempt to capture longer-range strain interactions caused by the point defects, in cells much smaller than would be required by a more conventional supercell (i.e. an $n \times n \times n$ supercell of the cubic bulk). \texttt{matscipy} \cite{matscipy} was used to generate the stacking fault and dislocation quadrupole structures.

\begin{table}[htb]
    \centering
    \caption[A list of configuration types in the InP dataset]{An alphabetized list of configuration types (categories of atomic structure) in the InP dataset, as well as information about the number of energy, force, and stress observations.}
    \scriptsize
    \begin{tabular}{@{}r|rrrr@{}}
\toprule
Configuration Type & $N_\text{Structures}$ & $N_\text{Atoms}$ & $N_\text{Forces}$ & $N_\text{Stresses}$ \\
\midrule
1000K\_PointDefect\_AL & 28 & 1,725 & 5,367 &  \\
300K\_NDSC\_Bulk & 50 & 1,457 & 4,416 &  \\
500K\_PointDefectNDSC & 261 & 3,389 & 10,191 &  \\
Accurate\_Bulk & 87 & 553 & 1,680 & 783 \\
Accurate\_Bulk\_MD & 11 & 81 & 264 &  \\
Accurate\_ZB\_Strain & 50 & 393 & 1,200 & 450 \\
Antisite & 235 & 10,625 & 31,968 &  \\
Bulk & 71 & 313 & 948 & 639 \\
Bulk\_Lowtemp\_MD & 16 & 961 & 3,072 &  \\
EOS & 104 & 621 & 1,872 & 936 \\
FlexCell\_PD\_Relax & 25 & 1,219 & 3,846 &  \\
GammaSurface\_MD & 448 & 19,789 & 59,649 &  \\
GammaSurface\_Relax & 144 & 5,073 & 15,264 &  \\
Interstitial & 229 & 14,821 & 44,655 &  \\
IsolatedAtom & 2 & 2 & 6 &  \\
KinkQuadrupole & 29 & 4,225 & 13,248 &  \\
PDMigrationCorrection & 26 & 1,602 & 4,992 &  \\
PD\_MD\_Bias & 9 & 286 & 948 &  \\
PD\_NEB\_Rattles & 31 & 1,949 & 6,039 &  \\
PD\_Rattles & 10 & 286 & 948 &  \\
PD\_Site\_Corr & 16 & 976 & 3,120 &  \\
PD\_Strain\_MD & 136 & 8,716 & 26,340 &  \\
PD\_Vac\_Antisite\_Rattles & 47 & 2,899 & 8,883 &  \\
QuadFormCorrection & 14 & 1,249 & 4,032 &  \\
QuadPD & 56 & 10,536 & 32,178 &  \\
Quadrupole\_Glide & 210 & 24,193 & 73,152 &  \\
Quadrupole\_Jog & 20 & 2,737 & 8,640 &  \\
Quadrupole\_MD & 105 & 17,761 & 53,856 &  \\
StackFaultISF\_PD & 49 & 4,617 & 14,133 &  \\
StackingFaultRattle & 149 & 11,953 & 35,928 &  \\
Strain\_MD & 300 & 1,761 & 5,292 &  \\
Vacancy & 311 & 12,171 & 36,699 &  \\
\midrule
Total & 3,279 & 168,939 & 512,826 & 2,808 \\
\bottomrule
\end{tabular}
    \label{tab:datasetlist}
\end{table}

Table \ref{tab:datasetlist} provides an overview of the contents of the dataset, partitioned by Configuration Type, which provides a concise label for the property or application a particular structure was designed to target. The table also includes information about the number of structures, atoms, and force and stress observations. Stress information is only included in a small subset of the dataset.

\subsection{DFT Parameterisation}
Data was generated using CASTEP 22.11 \cite{CASTEP}, using the RSCAN \cite{RSCAN} functional and on-the-fly ultrasoft pseudopotentials \cite{USPP1, USPP2} automatically generated by CASTEP. We use a 900 eV cutoff energy, and an SCF energy tolerance of $10^{-8}$ eV to ensure the accuracy of the quantum mechanical data. The K-point grids were generated through a Monkhorst-Pack scheme \cite{MonkhorstPack} in ASE \cite{ASE}, generating even grids with a density of at least 5.305 k-points per $\text{\AA}^{-1}$ and a grid offset of $\frac{1}{2 N} \text{\AA}^{-1}$, where $N$ is the size of the grid along a lattice vector. This is approximately equivalent to a \texttt{kpoints\_mp\_spacing} of around 0.03 $\text{\AA}^{-1}$ in the CASTEP \texttt{.cell} file.

\begin{table}
\caption{Comparison of lattice constant $a_0$, elastic constants $C_{11}$, $C_{12}$ and $C_{44}$, and band gap $E_b$ for LDA, PBE, and RSCAN exchange correlation functionals, compared with experiment. Experimental results are from (a) Nichols \cite{Nichols}, (b) Vasil'ev \cite{Vasilev}, (c) Hickernell \cite{Hickernell}, (d) Cunnel \cite{Cunnel}, and (e) Vurgaftman \cite{Vurgaftman}.}
\label{tbl:XC}

\begin{tabular}{l||rr|rrr}
\toprule
 & Expt. &  & LDA & PBE & RSCAN \\
\midrule
$a_0$ ($\AA$) & 5.856$^\text{a}$ & 5.87$^\text{b}$ & 5.82 & 5.96 & 5.9 \\
$C_{11}$ (GPa) & 101.1$^\text{a}$ & 102.2$^\text{c}$ & 100.53 & 86.74 & 97.98 \\
$C_{12}$ (GPa) & 56.1$^\text{a}$ & 57.6$^\text{c}$ & 56.58 & 45.6 & 53.31 \\
$C_{44}$ (GPa) & 45.6$^\text{a}$ & 46.0$^\text{c}$ & 45.56 & 41.73 & 46.3 \\
$E_b$ (eV) & 1.25$^\text{d}$ & 1.42$^\text{e}$ & 0.36 & 0.59 & 1.0 \\
\bottomrule
\end{tabular}
\end{table}

The RSCAN functional was chosen from the set of available functionals in CASTEP based on a test reproducing experimental elastic constants, and the band gap, as shown in Table \ref{tbl:XC}. The k-point density and cutoff energy were chosen based on a convergence test, where the energy and force errors were measured relative to a very high accuracy simulation on a rattled 8 atom cubic bulk structure. A 900 eV cutoff and a $6\times6\times6$ k-point grid is sufficient to converge the total energy to below 1 meV/Atom with a maximum force error of $\mathcal{O}(10 $meV$/\text{\AA})$, relative to a calculation with a 1500 eV cutoff energy and a $12\times12\times12$ grid. The choice of a $6\times6\times6$ grid was then converted into the sampling density $0.03\, \text{\AA}^{-1}$ for consistency across cells of differing dimensions.

\subsection{ACE \& MACE Parameterisation}
Atomic Cluster Expansion (ACE) \cite{ACE} and MACE \cite{MACE} models were trained on the InP dataset described above. The hyperparameters of the models (i.e. parameters which affect the model architecture, functional form, and/or number of learnable parameters) were chosen to achieve a high level of accuracy (i.e. Energy and Force RMSEs of $\mathrm{O}$(1 meV/Atom) and $\mathrm{O}$(10 meV/\AA) respectively), whilst minimising the evaluation cost of the models.

ACE is a linear model based on the ACE descriptor. The ACE model uses a 6 $\text{\AA}$ cutoff radius, a correlation order of 3 (meaning it encodes up to 4-body interactions), and a total degree of 18 (which means 18 2-body basis functions form the core of the basis). The model has a total of 15,006 basis functions, and was trained using the ``BLR" Bayesian regression solver from \texttt{ACEfit.jl} within the ACEpotentials code \cite{ACEpotentials}. A $p=4$ smoothness prior was used, as well as weights on the energy, force, and stress observations of 20 eV$^{-1}$, 1 (eV/\AA)$^{-1}$, and 5 (eV/\AA$^2$)$^{-1}$ respectively.

MACE is a Message-Passing Neural Network (MPNN) model, where vector messages encoding an atomic environment are passed to neighbouring atoms during inference. The MACE model used a 90\%:10\% train:validation split of the dataset, which was done on a per-configuration type basis (i.e. a random 10\% of each configuration type was used as validation data, and the rest used as training data). The model uses ``64x0e + 64x1o" features (meaning each message passing step passes 64 scalar and 64 vector equivariant messages from each atomic environment, to each connected environment), and was trained with a batch size of 16 configurations for 150 epochs. The model architecture leads to a total of 213,456 model weights. Stochastic Weight Averaging (SWA) \cite{SWA} was enabled from epoch 80, where the weights proposed by conventional gradient descent are averaged with the weights from previous iterations to improve the generalisability of the model. The model uses energy, force, and stress weights of 10 eV$^{-1}$, 1000 (eV/\AA)$^{-1}$, and 1000 (eV/\AA$^2$)$^{-1}$ initially, and then 1000 eV$^{-1}$, 10 (eV/\AA)$^{-1}$, and 1000 (eV/\AA$^2$)$^{-1}$ for the SWA portion of the training (i.e. the SWA portion of the training has greater bias towards energy accuracy, and less bias to force accuracy), with a fixed learning rate of 0.005.

\section{Results}
\label{sec:Results}
To make a case for the usefulness of the new InP potentials, it is not only important to validate them against direct DFT calculations, but also to compare with similar potentials from the literature. A summary of the performance of each model across a range of benchmarks is included in Tables \ref{tbl:Bulk} - \ref{tbl:Dataset}. Where appropriate, percentage errors relative to the DFT results are included.

\subsection{Bulk Properties}
We benchmark the models on the standard set of bulk properties, including the volume and lattice constant which minimise the potential energy, as well as elastic constants. Table \ref{tbl:Bulk} shows the results of the set of models applied to these quantities of interest. We see that all of the models give reasonable estimates of the lattice parameter, but that each of the previous literature models show maximum errors of $>10\%$ on the elastic constants predicted by RSCAN DFT.

We also compute an ``Equation of State" curve for both zincblende and wurtzite crystals, which measures the relative potential energy of the crystal as a function of the cell volume. To change the volume, we apply equal strain along each normal axis. The zero point for the energy scale is chosen as the minimum energy on the zincblende curve, which allows us to account for shifts in the definition of potential energy between models, whilst preserving the relative formation energy of the wurtzite crystal.

In addition to the Equation of State and other bulk properties, we also calculate a Phonon spectrum along the crystal momentum path measured by Borcherds et.\nobreak\,al.\nobreak\, \cite{Borcherds_1975}, and reproduce the data points from that publication for comparison with the set of potentials. We also compute the overall Density of States for Phonons, as an extra means of drawing comparisons between models.

\begin{figure}[ht]
    \centering
    \includegraphics[width=0.9\linewidth]{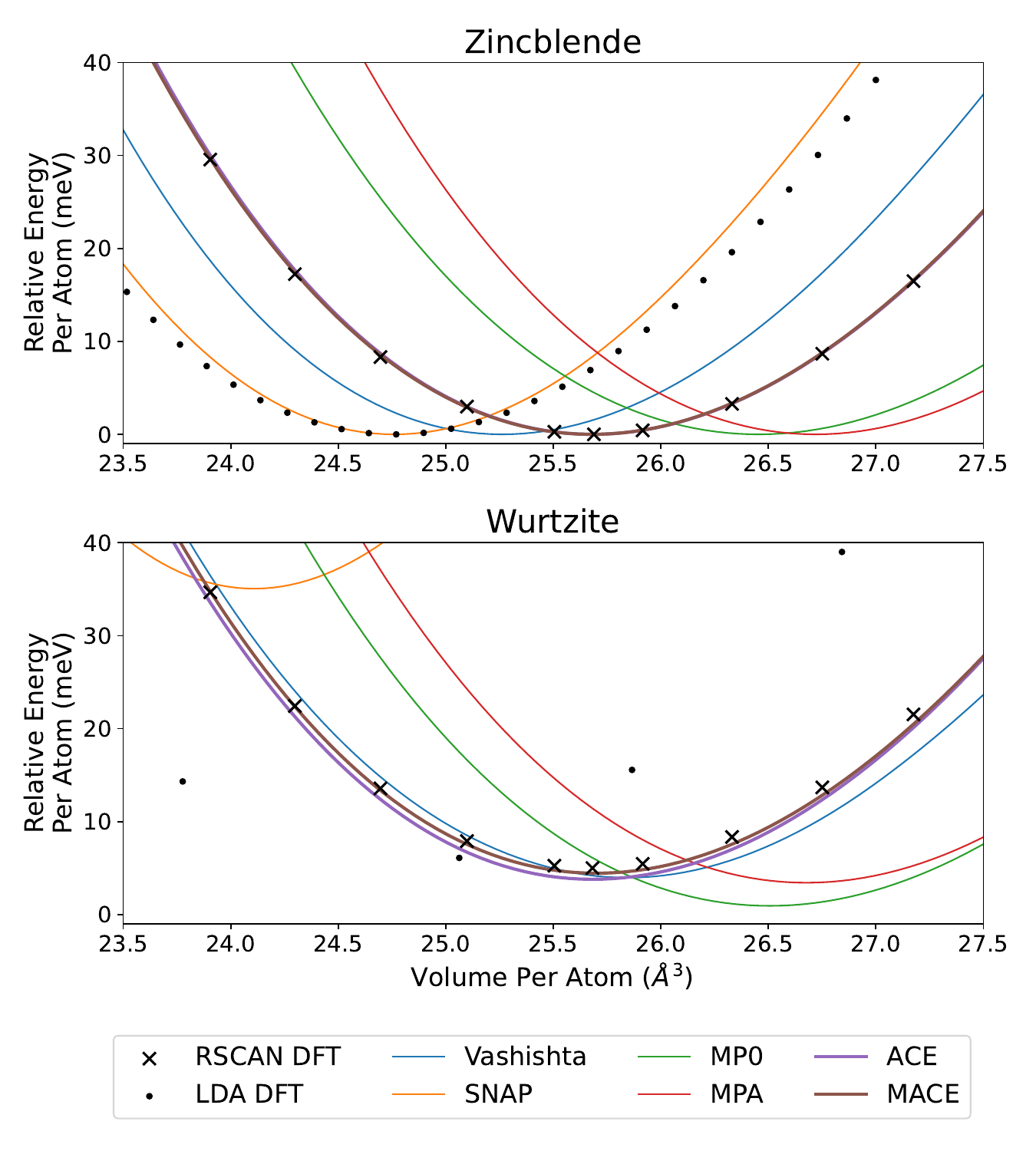}
    \caption[Equations of State predictions for InP potentials]{Equations of State (EOS) predictions for InP potentials on both zincblende (top panel) and wurtzite (bottom panel) crystal structures. The LDA DFT is taken from the SNAP training dataset \cite{SNAP}}
    \label{fig:EOS}
\end{figure}

\begin{figure}[ht]
    \centering
    \includegraphics[width=0.9\linewidth]{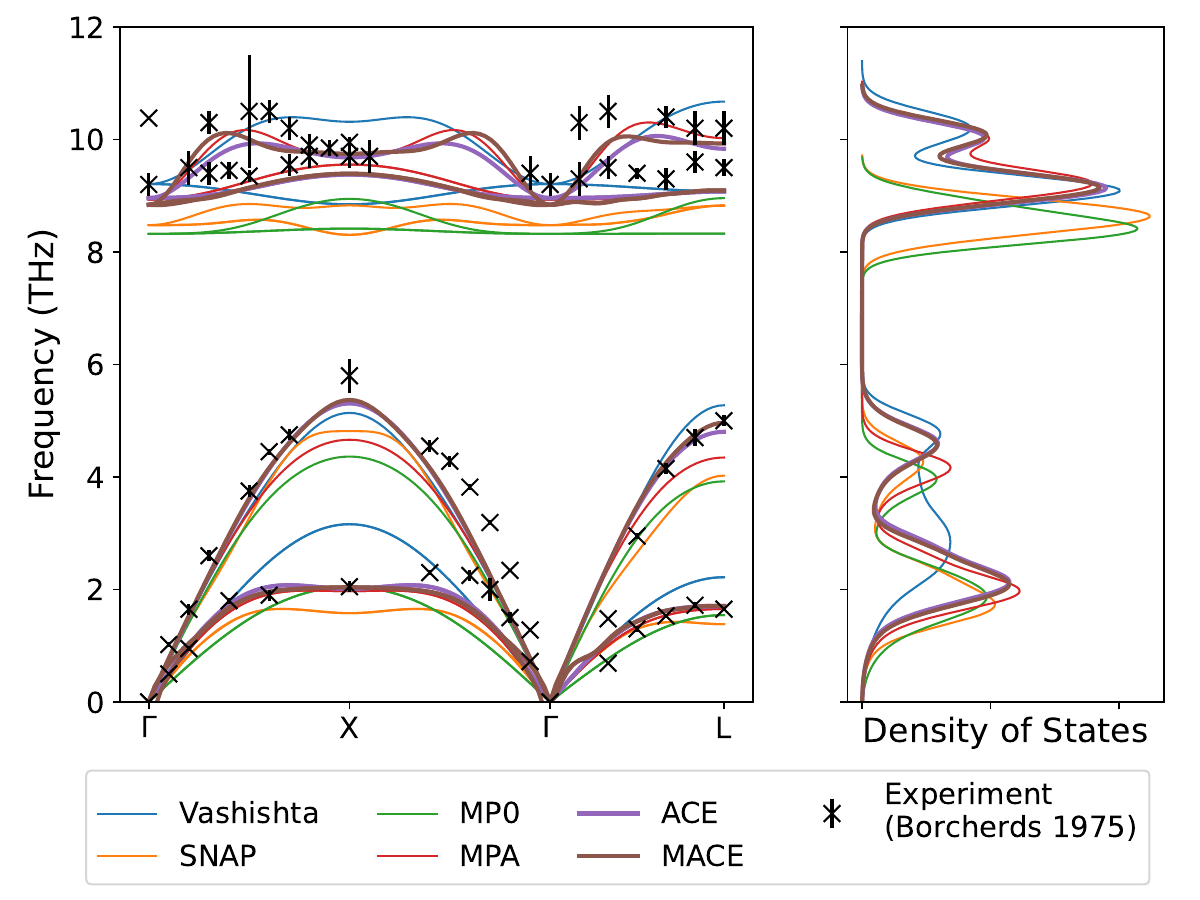}
    \caption[Phonon spectrum, compared with experimental observations from Borcherds et. al.]{Phonon spectrum, compared with experimental observations from Borcherds et.\nobreak\,al.\nobreak\,\cite{Borcherds_1975}. The left panel shows the phonon band structure along the $\Gamma X \Gamma L$ path used by Borcherds et.\nobreak\,al. The right panel shows a predicted density of states across the full Brillouin zone for each potential.}
    \label{fig:phonons}
\end{figure}

From Figure \ref{fig:EOS}, we see that there is some disagreement in the stable volume of both crystals between the models. As shown by the LDA DFT points (which were taken from the dataset used by the SNAP \cite{SNAP} model), this disagreement can in part be described by disagreements in the DFT functional and parameterisation (the MP0 and MPA models are based on PBE DFT from the Materials Project Dataset \cite{MaterialsProject}).  We see that the ACE and MACE models closely match the RSCAN DFT reference for both zincblende and wurtzite, the SNAP model matches the LDA reference well for zincblende, but incurs significant error on wurtzite. The Branicio Vashishta model fits the RSCAN DFT data well for wurtzite, but for zincblende predicts a stable volume which is between the LDA and RSCAN estimates. The MP0 and MPA models both overestimate the stable volume in comparison with both the LDA and RSCAN DFT, but the MPA model does predict the relative formation energy of wurtzite well in comparison to the DFT reference.

From Figure \ref{fig:phonons}, we see that the ACE and MACE closely match the experimentally observed frequencies of the acoustic branch, and provide a good approximation of the optical branch. The MPA model also gives reasonable predictions of both branches. The MP0, Vashishta, and SNAP potentials model some of the modes with an incorrect shape, and the MP0 and SNAP models underestimate the frequency of the optical branch. The MPA, ACE, and MACE models are also fairly self-consistent in the description of the Density of States.

\begin{table*}
\scriptsize
\caption{Comparison of Bulk Properties}
\label{tbl:Bulk}
\begin{tabular}{l||r|rrrr|rr}
\toprule
 & \begin{tabular}{r} RSCAN \\ DFT \end{tabular} & Vashishta & SNAP & MP0 & MPA & ACE & MACE \\
\midrule
\begin{tabular}{l}V$_0$ (Å$^3$)\end{tabular} & \begin{tabular}{r}25.68 \\ \end{tabular} & \begin{tabular}{r}25.26 \\ \textit{$-$2\%}\end{tabular} & \begin{tabular}{r}24.76 \\ \textit{$-$4\%}\end{tabular} & \begin{tabular}{r}26.45 \\ \textit{$+$3\%}\end{tabular} & \begin{tabular}{r}26.72 \\ \textit{$+$4\%}\end{tabular} & \begin{tabular}{r}25.68 \\ \textit{$+$0\%}\end{tabular} & \begin{tabular}{r}25.68 \\ \textit{$-$0\%}\end{tabular} \\
\begin{tabular}{l}a$_0$ (Å)\end{tabular} & \begin{tabular}{r}5.9 \\ \end{tabular} & \begin{tabular}{r}5.87 \\ \textit{$-$1\%}\end{tabular} & \begin{tabular}{r}5.83 \\ \textit{$-$1\%}\end{tabular} & \begin{tabular}{r}5.96 \\ \textit{$+$1\%}\end{tabular} & \begin{tabular}{r}5.98 \\ \textit{$+$1\%}\end{tabular} & \begin{tabular}{r}5.9 \\ \textit{$+$0\%}\end{tabular} & \begin{tabular}{r}5.9 \\ \textit{$-$0\%}\end{tabular} \\
\begin{tabular}{l}C$_{11}$ (GPa)\end{tabular} & \begin{tabular}{r}98.0 \\ \end{tabular} & \begin{tabular}{r}102.23 \\ \textit{$+$4\%}\end{tabular} & \begin{tabular}{r}113.73 \\ \textit{$+$16\%}\end{tabular} & \begin{tabular}{r}75.97 \\ \textit{$-$22\%}\end{tabular} & \begin{tabular}{r}92.52 \\ \textit{$-$6\%}\end{tabular} & \begin{tabular}{r}97.82 \\ \textit{$-$0\%}\end{tabular} & \begin{tabular}{r}98.21 \\ \textit{$+$0\%}\end{tabular} \\
\begin{tabular}{l}C$_{12}$ (GPa)\end{tabular} & \begin{tabular}{r}53.33 \\ \end{tabular} & \begin{tabular}{r}57.35 \\ \textit{$+$8\%}\end{tabular} & \begin{tabular}{r}70.94 \\ \textit{$+$33\%}\end{tabular} & \begin{tabular}{r}53.68 \\ \textit{$+$1\%}\end{tabular} & \begin{tabular}{r}56.11 \\ \textit{$+$5\%}\end{tabular} & \begin{tabular}{r}53.4 \\ \textit{$+$0\%}\end{tabular} & \begin{tabular}{r}52.56 \\ \textit{$-$1\%}\end{tabular} \\
\begin{tabular}{l}C$_{44}$ (GPa)\end{tabular} & \begin{tabular}{r}46.92 \\ \end{tabular} & \begin{tabular}{r}37.1 \\ \textit{$-$21\%}\end{tabular} & \begin{tabular}{r}48.47 \\ \textit{$+$3\%}\end{tabular} & \begin{tabular}{r}19.52 \\ \textit{$-$58\%}\end{tabular} & \begin{tabular}{r}38.58 \\ \textit{$-$18\%}\end{tabular} & \begin{tabular}{r}46.58 \\ \textit{$-$1\%}\end{tabular} & \begin{tabular}{r}46.53 \\ \textit{$-$1\%}\end{tabular} \\
\bottomrule
\end{tabular}
\end{table*}

\subsection{Point Defect Properties}
Point Defect formation energies are calculated using the approach of Freysoldt et.\nobreak \,\,al.\nobreak\,\,\cite{PDForm2}. We choose to use isolated atom energies as thermodynamic references when determining acceptable ranges for chemical potentials. The SNAP and Vashishta potentials were not trained on the more conventional elemental reference systems, for example elemental indium bulk, and so using a more conventional approach would create an unfair test for these models (as the point defect formation energy errors could become dominated by an energy error on the elemental reference, rather than an enery error on the point defect structure). We present formation energies calculated at the midpoint of the acceptable chemical potential ranges, thus
\begin{equation}
\begin{split}
    \mu_\text{In} &= \frac{1}{2}\left(\mu_\text{In}^\text{Iso} + \mu_\text{InP} - \mu_\text{P}^\text{Iso} \right) \\
    \mu_\text{P} &= \mu_\text{InP} - \mu_\text{In}
\end{split}
\end{equation}
where $\mu_\text{In}$ and $\mu_\text{P}$ are the chemical potentials, $\mu_\text{In}^\text{Iso}$ and $\mu_\text{P}^\text{Iso}$ are the energies of the isolated atom reference structures, and $\mu_\text{InP}$ is the energy of the InP zincblende primitive cell.

\begin{figure}[ht]
    \centering
    \includegraphics[width=0.9\linewidth]{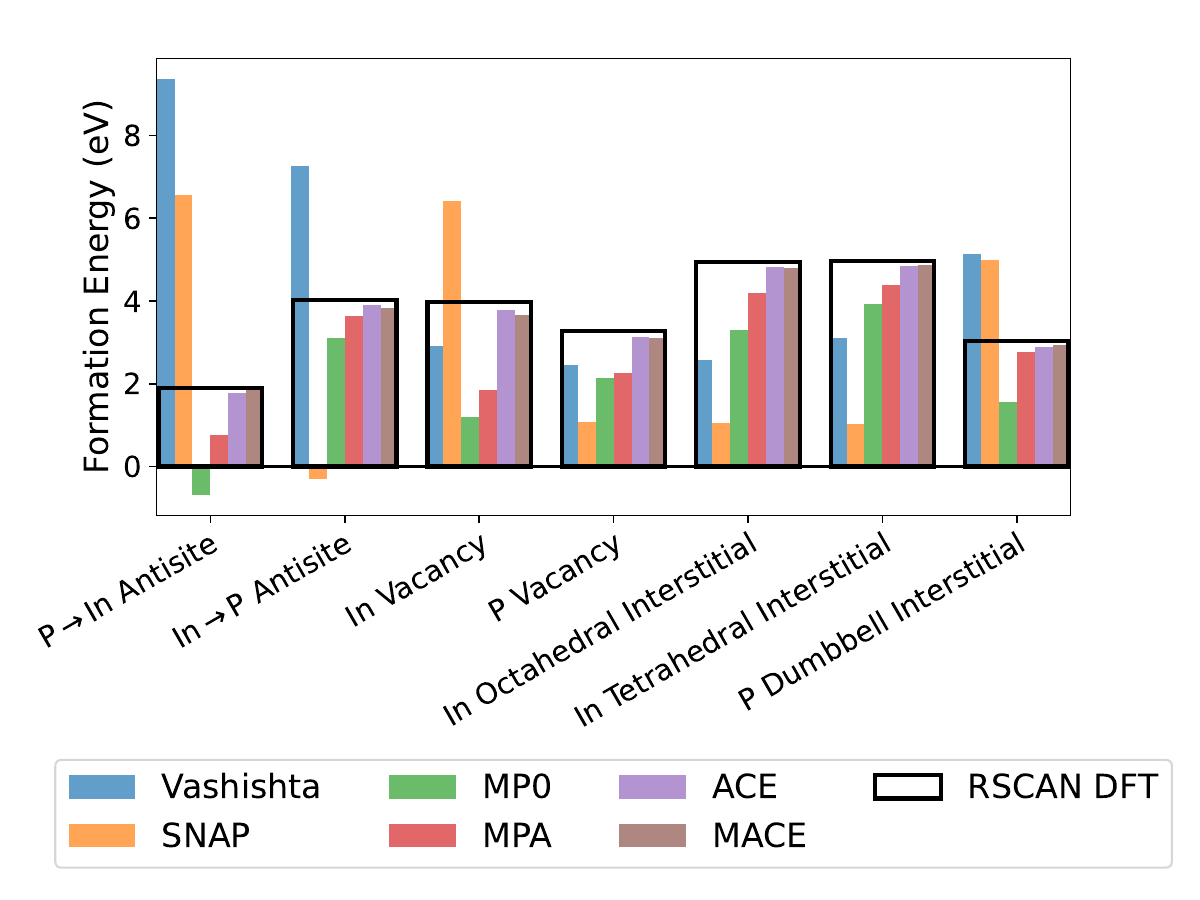}
    \caption[Comparison of Point Defect Formation Energies]{Comparison of Point Defect Formation Energies in zincblende for several InP potentials. The Indium Dumbbell and Phosphorus Octahedral and Tetrahedral Interstitials theoretically exist as additional defects, but preliminary testing found these defects were likely inaccessible thermodynamically.}
    \label{fig:PointDefect}
\end{figure}

\begin{figure*}[ht]
    \includegraphics[width=0.9\linewidth]{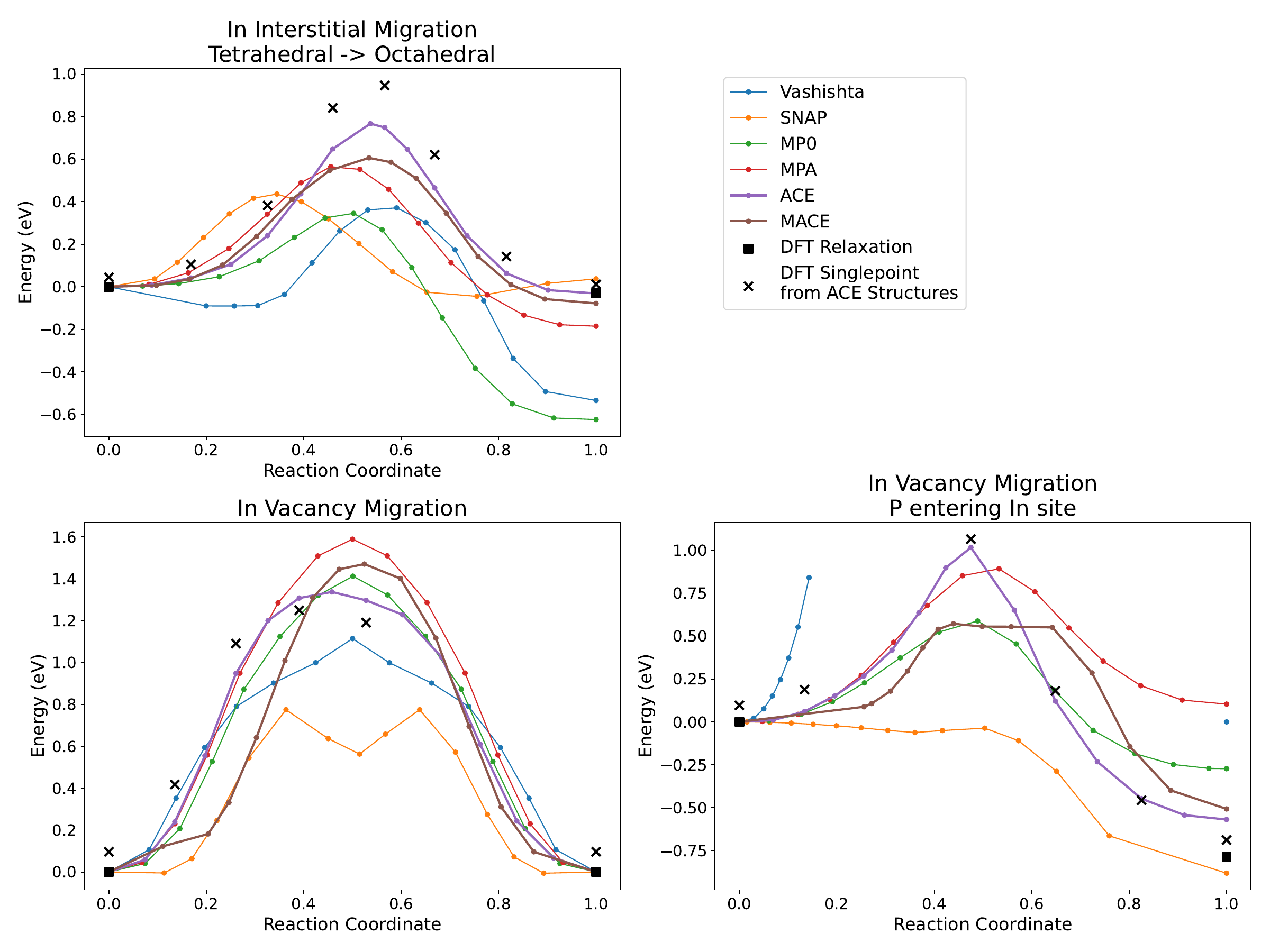}
    \caption[Indium Interstitial \& Vacancy Migration Barriers]{Indium Interstitial \& Vacancy Migration Barriers, as predicted by several potentials. The top left panel shows the Indium interstitial migration barrier between the tetrahedral and octahedral interstitial sites. The bottom left panel shows the Indium vacancy migration barrier. The bottom right panel shows the barrier for a Phosphorus atom to move into an Indium vacancy site, creating a Phosphorus vacancy and an In$\rightarrow$P antisite defect. The Vashishta model failed to converge to a meaningful barrier for the P atom entering the In Vacancy site, extremely large energies have been omitted from the plot.}
    \label{fig:In_Defect_Barriers}
\end{figure*}

\begin{figure*}[ht]
    \includegraphics[width=0.9\linewidth]{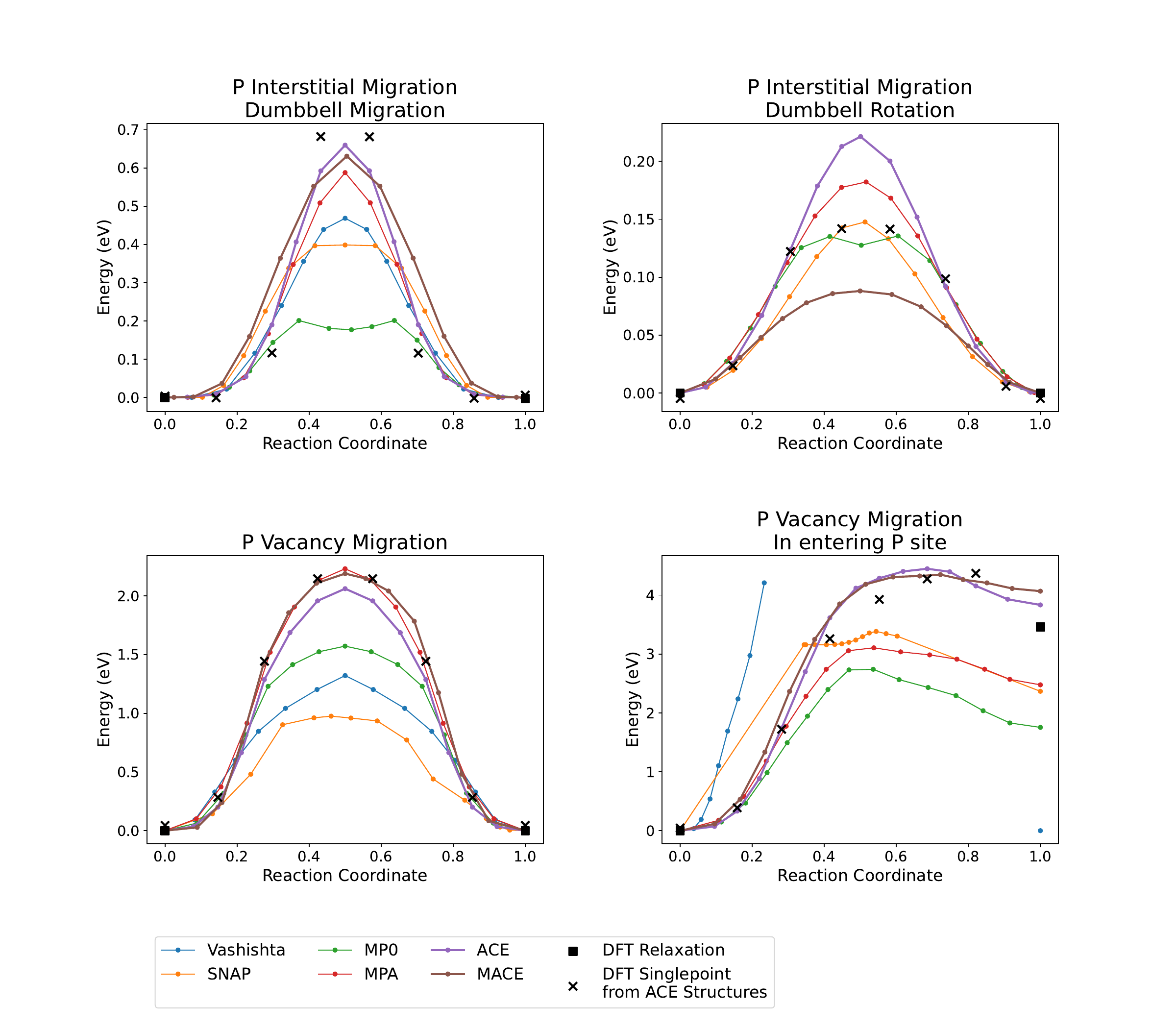}
    \caption[Phosphorus Interstitial \& Vacancy Migration Barriers]{Phosphorus Interstitial \& Vacancy Migration Barriers, as predicted by several potentials. The top left panel shows the Phosphorus dumbbell migration barrier. The top right panel shows the barrier for the dumbbell defect to rotate around the center of the dumbbell. The bottom left panel shows the barrier for a Phosphorus vacancy to migrate. The bottom right panel shows the barrier for an Indium atom to move into a Phosphorus vacancy site, creating an Indium vacancy and a P$\rightarrow$In antisite defect. The Vashishta model failed to converge to a meaningful barrier for the In atom entering the P Vacancy site, extremely large energies have been omitted from the plot.}
    \label{fig:P_Defect_Migration}
\end{figure*}

\begin{table*}
\scriptsize
\caption{Comparison of Point Defect Formation Energies (eV)}
\label{tbl:PointDefect}
\begin{tabular}{l||r|rrrr|rr}
\toprule
 & \begin{tabular}{r} RSCAN \\ DFT \end{tabular} & Vashishta & SNAP & MP0 & MPA & ACE & MACE \\
\midrule
\begin{tabular}{l} P$\rightarrow$In \\ Antisite \end{tabular} & \begin{tabular}{r}1.9 \\ \end{tabular} & \begin{tabular}{r}9.36 \\ \textit{$+$394\%}\end{tabular} & \begin{tabular}{r}6.57 \\ \textit{$+$247\%}\end{tabular} & \begin{tabular}{r}-0.68 \\ \textit{$-$136\%}\end{tabular} & \begin{tabular}{r}0.76 \\ \textit{$-$60\%}\end{tabular} & \begin{tabular}{r}1.78 \\ \textit{$-$6\%}\end{tabular} & \begin{tabular}{r}1.87 \\ \textit{$-$1\%}\end{tabular} \\
\begin{tabular}{l} In$\rightarrow$P \\ Antisite \end{tabular} & \begin{tabular}{r}4.02 \\ \end{tabular} & \begin{tabular}{r}7.26 \\ \textit{$+$81\%}\end{tabular} & \begin{tabular}{r}-0.29 \\ \textit{$-$107\%}\end{tabular} & \begin{tabular}{r}3.1 \\ \textit{$-$23\%}\end{tabular} & \begin{tabular}{r}3.65 \\ \textit{$-$9\%}\end{tabular} & \begin{tabular}{r}3.9 \\ \textit{$-$3\%}\end{tabular} & \begin{tabular}{r}3.84 \\ \textit{$-$5\%}\end{tabular} \\
\begin{tabular}{l} In Tetrahedral \\ Interstitial \end{tabular} & \begin{tabular}{r}4.98 \\ \end{tabular} & \begin{tabular}{r}3.11 \\ \textit{$-$37\%}\end{tabular} & \begin{tabular}{r}1.02 \\ \textit{$-$80\%}\end{tabular} & \begin{tabular}{r}3.93 \\ \textit{$-$21\%}\end{tabular} & \begin{tabular}{r}4.39 \\ \textit{$-$12\%}\end{tabular} & \begin{tabular}{r}4.86 \\ \textit{$-$2\%}\end{tabular} & \begin{tabular}{r}4.87 \\ \textit{$-$2\%}\end{tabular} \\
\begin{tabular}{l} In Octahedral \\ Interstitial \end{tabular} & \begin{tabular}{r}4.95 \\ \end{tabular} & \begin{tabular}{r}2.58 \\ \textit{$-$48\%}\end{tabular} & \begin{tabular}{r}1.06 \\ \textit{$-$79\%}\end{tabular} & \begin{tabular}{r}3.3 \\ \textit{$-$33\%}\end{tabular} & \begin{tabular}{r}4.21 \\ \textit{$-$15\%}\end{tabular} & \begin{tabular}{r}4.82 \\ \textit{$-$2\%}\end{tabular} & \begin{tabular}{r}4.8 \\ \textit{$-$3\%}\end{tabular} \\
\begin{tabular}{l} P Dumbbell \\ Interstitial \end{tabular} & \begin{tabular}{r}3.04 \\ \end{tabular} & \begin{tabular}{r}5.15 \\ \textit{$+$70\%}\end{tabular} & \begin{tabular}{r}5.0 \\ \textit{$+$65\%}\end{tabular} & \begin{tabular}{r}1.57 \\ \textit{$-$48\%}\end{tabular} & \begin{tabular}{r}2.76 \\ \textit{$-$9\%}\end{tabular} & \begin{tabular}{r}2.89 \\ \textit{$-$5\%}\end{tabular} & \begin{tabular}{r}2.94 \\ \textit{$-$3\%}\end{tabular} \\
\begin{tabular}{l}In Vacancy\end{tabular} & \begin{tabular}{r}3.97 \\ \end{tabular} & \begin{tabular}{r}2.91 \\ \textit{$-$27\%}\end{tabular} & \begin{tabular}{r}6.41 \\ \textit{$+$62\%}\end{tabular} & \begin{tabular}{r}1.19 \\ \textit{$-$70\%}\end{tabular} & \begin{tabular}{r}1.86 \\ \textit{$-$53\%}\end{tabular} & \begin{tabular}{r}3.78 \\ \textit{$-$5\%}\end{tabular} & \begin{tabular}{r}3.66 \\ \textit{$-$8\%}\end{tabular} \\
\begin{tabular}{l}P Vacancy\end{tabular} & \begin{tabular}{r}3.27 \\ \end{tabular} & \begin{tabular}{r}2.46 \\ \textit{$-$25\%}\end{tabular} & \begin{tabular}{r}1.09 \\ \textit{$-$67\%}\end{tabular} & \begin{tabular}{r}2.15 \\ \textit{$-$34\%}\end{tabular} & \begin{tabular}{r}2.27 \\ \textit{$-$31\%}\end{tabular} & \begin{tabular}{r}3.13 \\ \textit{$-$4\%}\end{tabular} & \begin{tabular}{r}3.1 \\ \textit{$-$5\%}\end{tabular} \\
\bottomrule
\end{tabular}
\end{table*}

\begin{table*}
\scriptsize
\caption{Comparison of Point Defect Max Positional Error ($\text{\AA}$)}
\label{tbl:PointDefectPos}
\begin{tabular}{l||r|rrrr|rr}
\toprule
 & \begin{tabular}{r} RSCAN \\ DFT \end{tabular} & Vashishta & SNAP & MP0 & MPA & ACE & MACE \\
\midrule
\begin{tabular}{l} P$\rightarrow$In \\ Antisite \end{tabular} & \begin{tabular}{r}- \\ \end{tabular} & \begin{tabular}{r}1.08 \\ \end{tabular} & \begin{tabular}{r}0.08 \\ \end{tabular} & \begin{tabular}{r}0.15 \\ \end{tabular} & \begin{tabular}{r}0.18 \\ \end{tabular} & \begin{tabular}{r}0.00 \\ \end{tabular} & \begin{tabular}{r}0.00 \\ \end{tabular} \\
\begin{tabular}{l} In$\rightarrow$P \\ Antisite \end{tabular} & \begin{tabular}{r}- \\ \end{tabular} & \begin{tabular}{r}1.10 \\ \end{tabular} & \begin{tabular}{r}0.10 \\ \end{tabular} & \begin{tabular}{r}0.05 \\ \end{tabular} & \begin{tabular}{r}0.01 \\ \end{tabular} & \begin{tabular}{r}0.00 \\ \end{tabular} & \begin{tabular}{r}0.00 \\ \end{tabular} \\
\begin{tabular}{l} In Tetrahedral \\ Interstitial \end{tabular} & \begin{tabular}{r}- \\ \end{tabular} & \begin{tabular}{r}0.15 \\ \end{tabular} & \begin{tabular}{r}0.11 \\ \end{tabular} & \begin{tabular}{r}0.03 \\ \end{tabular} & \begin{tabular}{r}0.03 \\ \end{tabular} & \begin{tabular}{r}0.01 \\ \end{tabular} & \begin{tabular}{r}0.02 \\ \end{tabular} \\
\begin{tabular}{l} In Octahedral \\ Interstitial \end{tabular} & \begin{tabular}{r}- \\ \end{tabular} & \begin{tabular}{r}0.20 \\ \end{tabular} & \begin{tabular}{r}0.05 \\ \end{tabular} & \begin{tabular}{r}0.07 \\ \end{tabular} & \begin{tabular}{r}0.07 \\ \end{tabular} & \begin{tabular}{r}0.01 \\ \end{tabular} & \begin{tabular}{r}0.01 \\ \end{tabular} \\
\begin{tabular}{l} P Dumbbell \\ Interstitial \end{tabular} & \begin{tabular}{r}- \\ \end{tabular} & \begin{tabular}{r}0.54 \\ \end{tabular} & \begin{tabular}{r}0.10 \\ \end{tabular} & \begin{tabular}{r}0.11 \\ \end{tabular} & \begin{tabular}{r}0.06 \\ \end{tabular} & \begin{tabular}{r}0.02 \\ \end{tabular} & \begin{tabular}{r}0.25 \\ \end{tabular} \\
\begin{tabular}{l}In Vacancy\end{tabular} & \begin{tabular}{r}- \\ \end{tabular} & \begin{tabular}{r}0.42 \\ \end{tabular} & \begin{tabular}{r}0.13 \\ \end{tabular} & \begin{tabular}{r}0.08 \\ \end{tabular} & \begin{tabular}{r}0.16 \\ \end{tabular} & \begin{tabular}{r}0.07 \\ \end{tabular} & \begin{tabular}{r}0.27 \\ \end{tabular} \\
\begin{tabular}{l}P Vacancy\end{tabular} & \begin{tabular}{r}- \\ \end{tabular} & \begin{tabular}{r}0.56 \\ \end{tabular} & \begin{tabular}{r}0.62 \\ \end{tabular} & \begin{tabular}{r}0.25 \\ \end{tabular} & \begin{tabular}{r}0.28 \\ \end{tabular} & \begin{tabular}{r}0.01 \\ \end{tabular} & \begin{tabular}{r}0.01 \\ \end{tabular} \\
\bottomrule
\end{tabular}
\end{table*}

Figure \ref{fig:PointDefect} shows the formation energy of various non-stoichiometric point defects in zincblende InP. Table \ref{tbl:PointDefect} also tabulates this information. We see that the ACE and MACE models are the only potentials to accurately reproduce the DFT reference to within a few percent, and that the Vashishta and SNAP models are particularly erroneous. Table \ref{tbl:PointDefectPos} shows the maximum error on atomic positions 
\begin{equation}
    \max_i \left|\bm{r}_i - \bm{r}_i^\text{ref} \right|
\end{equation}
relative to the DFT-relaxed structure. The results show that, aside from the Vashishta classical potential, the models generate relaxed structures which are extremely close to the DFT minimum.

The energy barriers for point defect migration were generated using the Nudged Elastic Band (NEB) approach as implemented in ASE \cite{ASE}. The DFT NEB was not performed due to the large computational expense of the calculation. However, the endpoints of each migration were fully relaxed with DFT to benchmark accuracy in the relative energies of the initial and final states, and static DFT energy evaluations were performed on a subset of structures from the ACE trajectory in order to provide a rough benchmark of the accuracy of the migration path. The ``reaction coordinate" used for the x-axis on the migration plots is the integrated displacement between images, which is then normalised on [0, 1].

Figures \ref{fig:In_Defect_Barriers} and \ref{fig:P_Defect_Migration} show predicted migration barriers for interstitial and vacancy defects respectively. Generally, we see the ACE, MACE, and MPA models capturing the correct shape of the barrier in most cases, and although the estimates of the barriers are less accurate than energy errors in other benchmarks, the rough scale is correct with these models.

\subsection{Stacking Faults}
Stacking fault energies are measured as the change in energy density as one crystal is displaced relative to another, due to changes in the interfaces. To achieve this, we first construct a periodic crystal of InP such that the surface normal vector is aligned along the z axis, and the displacement direction is parallel with the y axis. We can then apply the required displacements by changing the xz component of the periodic cell to create a single interface at the top and bottom of the cell. We choose to use the convention where we relax the structures with constraints such that atoms are only free to move in the z direction.

\begin{figure*}[htb]
    \centering
    \includegraphics[width=0.7\linewidth]{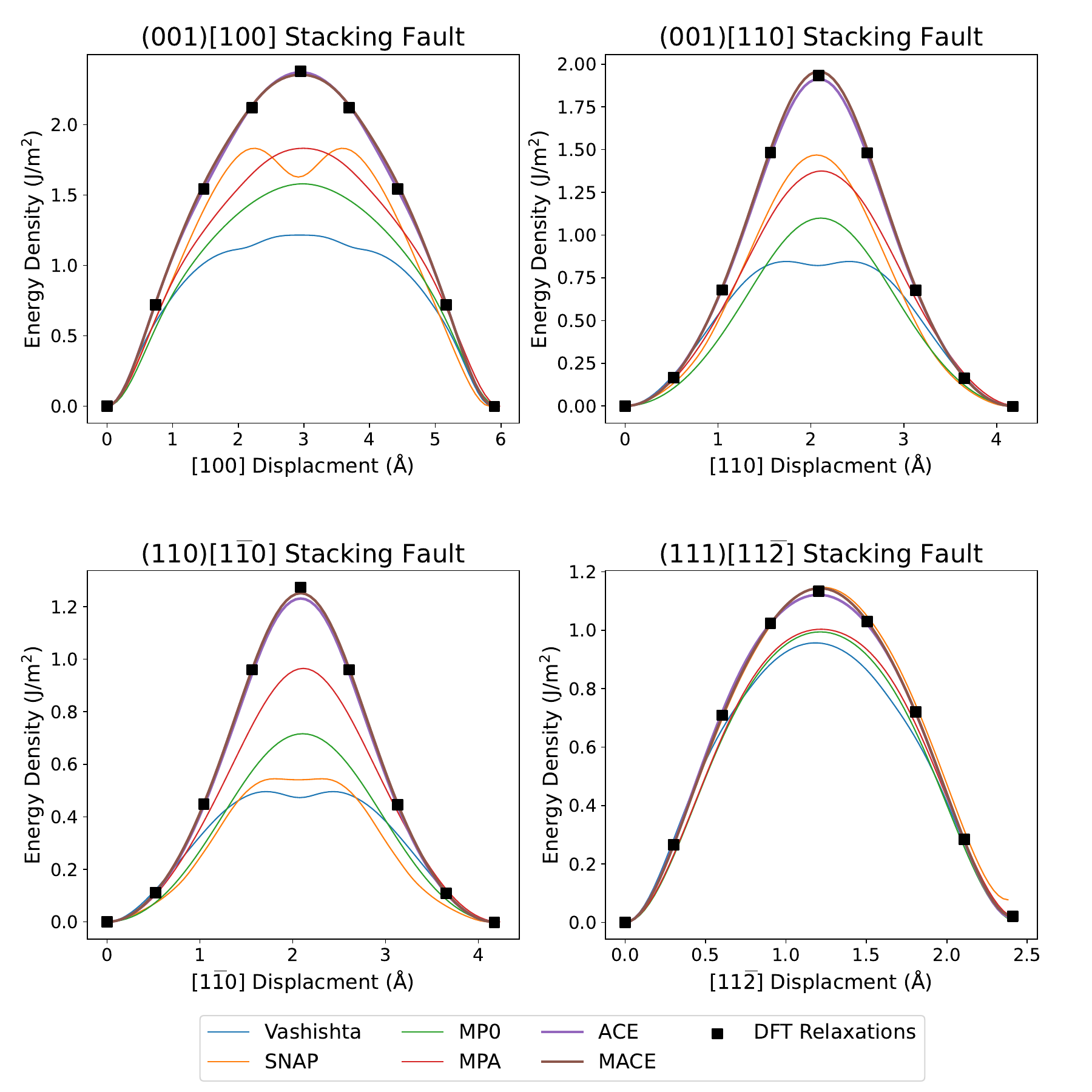}
    \caption[Predicted stacking fault barriers]{Comparison of predicted stacking fault barriers for several InP potentials. The four larger panels show the predicted stacking fault energy curve as a function of the displacement of the upper surface, for four different stacking faults. The twelve smaller panels show atomic structures at three points along the stacking fault line, for each of the four stacking faults. Only the (111)[11-2] fault results in a stable intrinsic stacking fault structure.}
    \label{fig:StackFault}
\end{figure*}

The $[11\bar{2}]$ intrinsic stacking fault (ISF) energy is the formation of the stable $[11\bar{2}
]$ stacking fault, which in the $[11\bar{2}]$ panel of Figure \ref{fig:StackFault} corresponds with the energy of the final point on each curve.

Figure \ref{fig:StackFault} shows the predicted stacking fault energy curves for a set of simple fault planes and directions. We again see that the ACE and MACE models reproduce the DFT reference extremely closely. The MP0 and MPA models generally show reasonably correct shapes of the curves, though both models generally underestimate the peak fault energy. The SNAP and Vashishta models show incorrect behaviour in the shape of the curves, as well as incorrect peak energy in all but the $(111)[11\bar{2}]$ fault, where the SNAP model is able to match the DFT reasonably closely.

\begin{table*}
\scriptsize
\caption{Comparison of Stacking Fault Energies (J/m$^2$)}
\label{tbl:StackFault}
\begin{tabular}{l||r|rrrr|rr}
\toprule
 & \begin{tabular}{r} RSCAN \\ DFT \end{tabular} & Vashishta & SNAP & MP0 & MPA & ACE & MACE \\
\midrule
\begin{tabular}{l}(001)[100]\end{tabular} & \begin{tabular}{r}2.38 \\ \end{tabular} & \begin{tabular}{r}1.22 \\ \textit{$-$49\%}\end{tabular} & \begin{tabular}{r}1.83 \\ \textit{$-$23\%}\end{tabular} & \begin{tabular}{r}1.58 \\ \textit{$-$34\%}\end{tabular} & \begin{tabular}{r}1.83 \\ \textit{$-$23\%}\end{tabular} & \begin{tabular}{r}2.37 \\ \textit{$-$0\%}\end{tabular} & \begin{tabular}{r}2.35 \\ \textit{$-$1\%}\end{tabular} \\
\begin{tabular}{l}(001)[110]\end{tabular} & \begin{tabular}{r}1.93 \\ \end{tabular} & \begin{tabular}{r}0.85 \\ \textit{$-$56\%}\end{tabular} & \begin{tabular}{r}1.47 \\ \textit{$-$24\%}\end{tabular} & \begin{tabular}{r}1.1 \\ \textit{$-$43\%}\end{tabular} & \begin{tabular}{r}1.37 \\ \textit{$-$29\%}\end{tabular} & \begin{tabular}{r}1.91 \\ \textit{$-$1\%}\end{tabular} & \begin{tabular}{r}1.96 \\ \textit{$+$1\%}\end{tabular} \\
\begin{tabular}{l}(110)[1$\bar{1}$0]\end{tabular} & \begin{tabular}{r}1.27 \\ \end{tabular} & \begin{tabular}{r}0.5 \\ \textit{$-$61\%}\end{tabular} & \begin{tabular}{r}0.54 \\ \textit{$-$57\%}\end{tabular} & \begin{tabular}{r}0.72 \\ \textit{$-$44\%}\end{tabular} & \begin{tabular}{r}0.96 \\ \textit{$-$24\%}\end{tabular} & \begin{tabular}{r}1.23 \\ \textit{$-$3\%}\end{tabular} & \begin{tabular}{r}1.25 \\ \textit{$-$2\%}\end{tabular} \\
\begin{tabular}{l}(111)[11$\bar{2}$]\end{tabular} & \begin{tabular}{r}1.13 \\ \end{tabular} & \begin{tabular}{r}0.96 \\ \textit{$-$16\%}\end{tabular} & \begin{tabular}{r}1.15 \\ \textit{$+$1\%}\end{tabular} & \begin{tabular}{r}0.99 \\ \textit{$-$12\%}\end{tabular} & \begin{tabular}{r}1.0 \\ \textit{$-$11\%}\end{tabular} & \begin{tabular}{r}1.12 \\ \textit{$-$1\%}\end{tabular} & \begin{tabular}{r}1.14 \\ \textit{$+$1\%}\end{tabular} \\
\begin{tabular}{l} (111)[11$\bar{2}$] \\ISF (mJ/m$^2$) \end{tabular} & \begin{tabular}{r}21.24 \\ \end{tabular} & \begin{tabular}{r}28.28 \\ \textit{$+$33\%}\end{tabular} & \begin{tabular}{r}77.24 \\ \textit{$+$264\%}\end{tabular} & \begin{tabular}{r}4.82 \\ \textit{$-$77\%}\end{tabular} & \begin{tabular}{r}14.32 \\ \textit{$-$33\%}\end{tabular} & \begin{tabular}{r}15.03 \\ \textit{$-$29\%}\end{tabular} & \begin{tabular}{r}19.43 \\ \textit{$-$9\%}\end{tabular} \\
\bottomrule
\end{tabular}
\end{table*}

Table \ref{tbl:StackFault} shows the predicted stacking fault barrier density (i.e. the energy density difference between the maximum in the stacking fault curve and bulk), as well as the predicted ISF energy of the $(111)[11\bar{2}]$ fault. We again see that the ACE and MACE models are able to reproduce the DFT almost exactly, where the other models show $>10\%$ error (aside from the SNAP model on the $(111)[11\bar{2}]$ fault, which is acceptably accurate). The ISF energy is an extremely challenging benchmark, as the observed errors correspond to only a few meV/Atom of error on the total energy of the stacking fault structure, which is the same order of magnitude of errors expected by MLIPs.

\subsection{Dislocation Quadrupoles}
Dislocation Quadrupoles are a means of generating periodic atomic structures which contain dislocation cores. A periodic system is required to have zero net Burgers vector, and so the quadrupole cells contain two dislocation cores of opposite burgers vector, arranged such that each core is equidistant to four cores of opposite burgers vector. 

Such structures are unphysical due to the extremely high dislocation density, but such cells are extremely useful for training and validating dislocations at a scale which is feasible with DFT directly. Here we only test on quadrupoles of the partial dislocations, as they are stable under relaxation in cells small enough to be directly relaxed by DFT at a reasonable computational cost - similar cells containing perfect Screw or $60^\circ$ dislocation cores experience such a large dislocation-dislocation interaction that relaxation of the structure causes the opposing burgers vectors to annihilate, resulting in a bulk structure. 

To benchmark the effectiveness of each potential at modeling dislocation systems, we compare quadrupole structures relaxed using each potential, and measure the formation energy of the quadrupole ($E_\mathrm{Quad} - E_\mathrm{Bulk}$), and the RMSE error in fractional positions relative to RSCAN DFT ($\left| \mathbf{r}_{i; \text{Pot}} - \mathbf{r}_{i; \text{DFT}} \right|$). Fractional coordinates are used so we are able to compare using the lattice parameter predicted by each potential.

\begin{table*}
\scriptsize
\caption{Comparison of Dislocation Quadrupole Properties}
\label{tbl:Disloc}
\begin{tabular}{l||r|rrrr|rr}
\toprule
 & \begin{tabular}{r} RSCAN \\ DFT \end{tabular} & Vashishta & SNAP & MP0 & MPA & ACE & MACE \\
\midrule
\begin{tabular}{l} $30^\circ$ Partial \\ Formation Energy (eV) \end{tabular} & \begin{tabular}{r}3.79 \\ \end{tabular} & \begin{tabular}{r}2.19 \\ \textit{$-$42\%}\end{tabular} & \begin{tabular}{r}1.15 \\ \textit{$-$70\%}\end{tabular} & \begin{tabular}{r}2.08 \\ \textit{$-$45\%}\end{tabular} & \begin{tabular}{r}3.09 \\ \textit{$-$18\%}\end{tabular} & \begin{tabular}{r}3.66 \\ \textit{$-$4\%}\end{tabular} & \begin{tabular}{r}3.72 \\ \textit{$-$2\%}\end{tabular} \\
\begin{tabular}{l} $30^\circ$ Partial RMSE on \\ scaled positions ($10^{-2}$) \end{tabular} & \begin{tabular}{r}- \\ \end{tabular} & \begin{tabular}{r}6.04 \\ \end{tabular} & \begin{tabular}{r}7.21 \\ \end{tabular} & \begin{tabular}{r}1.28 \\ \end{tabular} & \begin{tabular}{r}0.72 \\ \end{tabular} & \begin{tabular}{r}0.44 \\ \end{tabular} & \begin{tabular}{r}0.53 \\ \end{tabular} \\
\begin{tabular}{l} $30^\circ$ Max Positional \\ Error ($\text{\AA}$) \end{tabular} & \begin{tabular}{r}- \\ \end{tabular} & \begin{tabular}{r}1.11 \\ \end{tabular} & \begin{tabular}{r}1.42 \\ \end{tabular} & \begin{tabular}{r}0.31 \\ \end{tabular} & \begin{tabular}{r}0.35 \\ \end{tabular} & \begin{tabular}{r}0.11 \\ \end{tabular} & \begin{tabular}{r}0.13 \\ \end{tabular} \\
\begin{tabular}{l} $90^\circ$ Partial \\ Formation Energy (eV) \end{tabular} & \begin{tabular}{r}4.41 \\ \end{tabular} & \begin{tabular}{r}3.75 \\ \textit{$-$15\%}\end{tabular} & \begin{tabular}{r}- \\ \end{tabular} & \begin{tabular}{r}2.19 \\ \textit{$-$50\%}\end{tabular} & \begin{tabular}{r}3.64 \\ \textit{$-$18\%}\end{tabular} & \begin{tabular}{r}4.29 \\ \textit{$-$3\%}\end{tabular} & \begin{tabular}{r}4.24 \\ \textit{$-$4\%}\end{tabular} \\
\begin{tabular}{l} $90^\circ$ Partial RMSE on \\ scaled positions ($10^{-2}$) \end{tabular} & \begin{tabular}{r}- \\ \end{tabular} & \begin{tabular}{r}8.26 \\ \end{tabular} & \begin{tabular}{r}- \\ \end{tabular} & \begin{tabular}{r}1.39 \\ \end{tabular} & \begin{tabular}{r}1.47 \\ \end{tabular} & \begin{tabular}{r}0.52 \\ \end{tabular} & \begin{tabular}{r}0.37 \\ \end{tabular} \\
\begin{tabular}{l} $90^\circ$ Max Positional \\ Error ($\text{\AA}$) \end{tabular} & \begin{tabular}{r}- \\ \end{tabular} & \begin{tabular}{r}1.51 \\ \end{tabular} & \begin{tabular}{r}- \\ \end{tabular} & \begin{tabular}{r}0.37 \\ \end{tabular} & \begin{tabular}{r}0.35 \\ \end{tabular} & \begin{tabular}{r}0.07 \\ \end{tabular} & \begin{tabular}{r}0.07 \\ \end{tabular} \\
\bottomrule
\end{tabular}
\end{table*}

Table \ref{tbl:Disloc} shows the predicted quadrupole formation energies, and the error on the fractional coordinates, for each potential. We see again that ACE and MACE are very accurate to the DFT quadrupole structure, but also that MP0 and particularly MPA give reasonable predictions.

We also measure the energy barriers associated with the dissociation length of the quadrupole structure changing by one glide distance (+$3.55\text{\AA}$). Similarly to the point defect migration benchmark, here we only use DFT to relax the endpoints of the barriers, and to evaluate the energies of a subset of the ACE trajectory.
\begin{figure}[htb]
\centering
    \includegraphics[width=\linewidth]{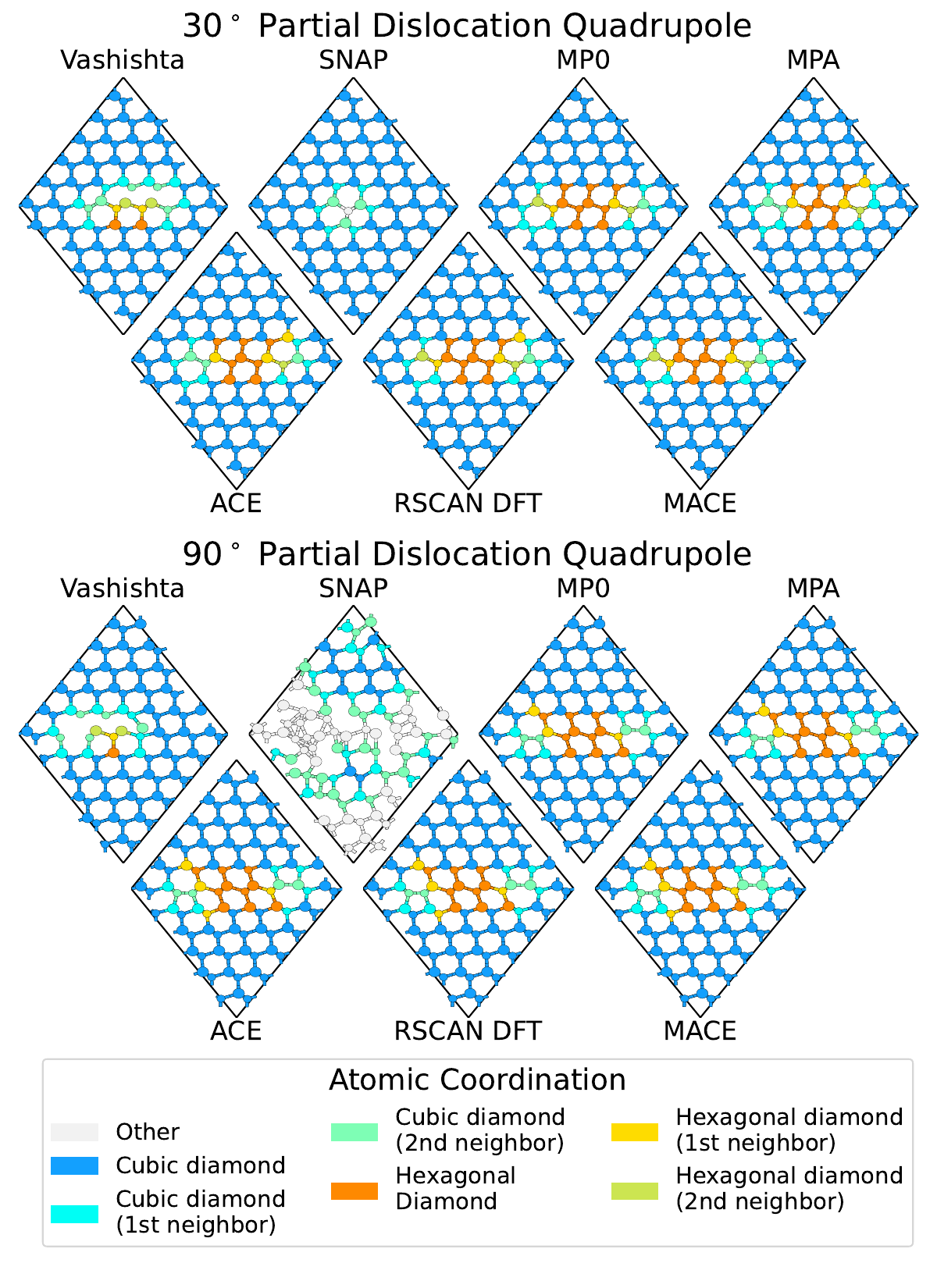}

    \caption[Comparison of relaxed dislocation quadrupole structures]{Comparison of relaxed dislocation quadrupole structures for several InP potentials. The top set of panels correspond to the 30$^\circ$ partial dislocation, and the bottom set show the 90$^\circ$ partial dislocation. Atom colours are generated via Ovito \cite{Ovito} \texttt{IdentifyDiamondModifier}, which is based on a neighbour analysis. Bonds are drawn with a 3Å cutoff. The SNAP potential did not converge the $90^\circ$ quadrupole in 400 LBFGS iterations, the structure shown is the result of the 400th iteration.}
    \label{fig:QuadFormation}
\end{figure}

\begin{figure*}[htb]
\centering
    \includegraphics[width=0.8\linewidth]{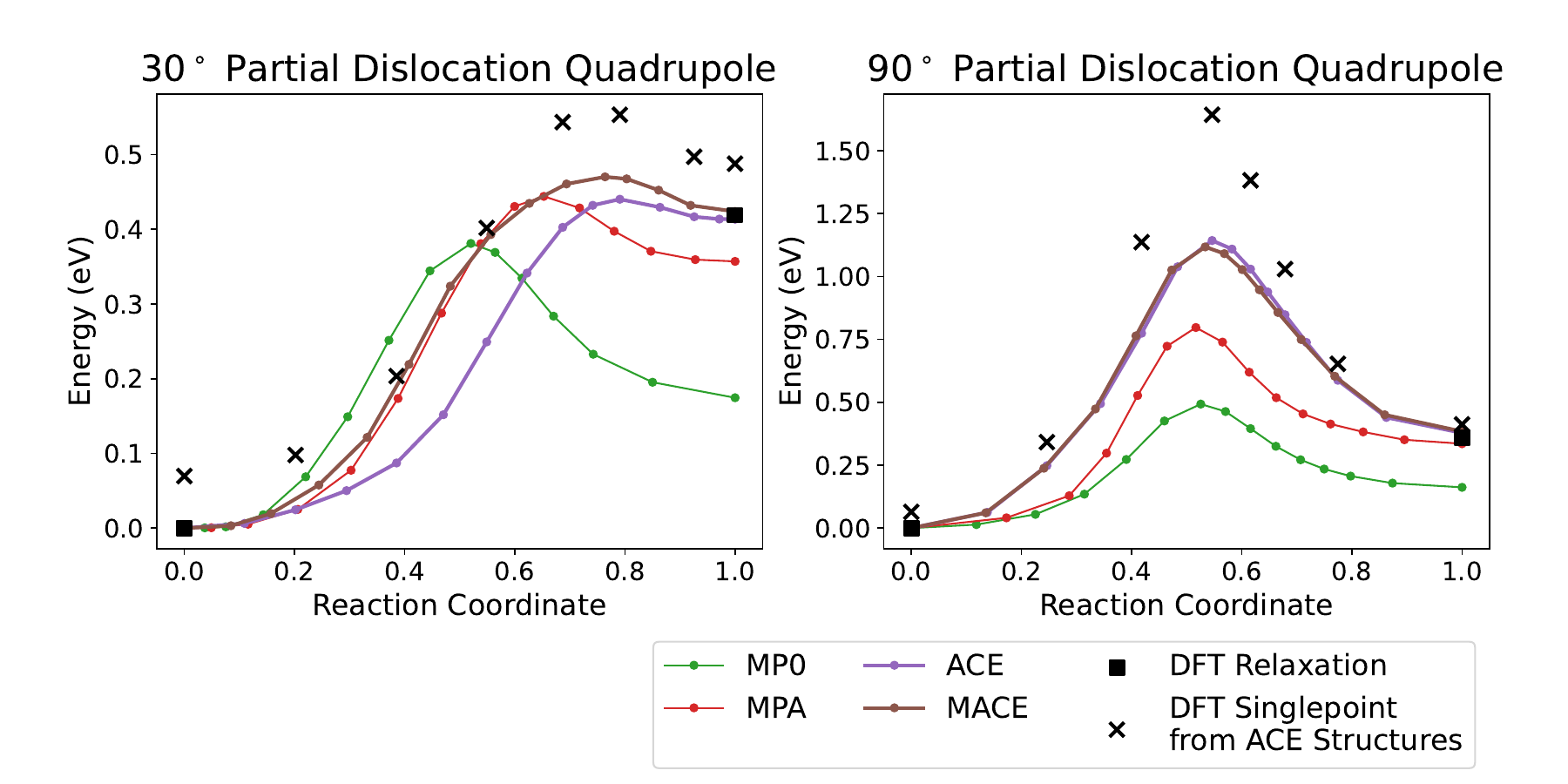}

    \caption[Migration Barriers for a partial core glide event in the quadrupole]{Migration Barriers for a partial core glide event in the quadrupole. The top panel shows the predicted barrier for a glide of a 30$^\circ$ partial dislocation within a quadrupole structure. The bottom panel shows the predicted barrier for a glide of a 90$^\circ$ partial dislocation within a quadrupole structure.}
    \label{fig:QuadGlide}
\end{figure*}

Figure \ref{fig:QuadFormation} shows a 2D projection of the quadrupole structures as relaxed by each model, where the view is the plane normal to the dislocation line direction. We see visually and via Common Neighbour Analysis that the MP0, MPA, ACE, and MACE models relax to structures which are very similar to the DFT relaxed structure. The Vashishta potential clearly produces stable but incorrect quadrupole structures, where the separation of the two partial dislocations has partially collapsed. The SNAP model almost entirely collapses the $30^\circ$ partial dislocation quadrupole structure. The $90^\circ$ partial quadrupole structure is likely too extrapolative for the SNAP model, as when a relaxation was attempted an LBFGS minimisation struggled to converge, and produced a structure where the conventional crystal pattern in this projection had been clearly disrupted.

Figure \ref{fig:QuadGlide} shows migration barriers for a glide event in a partial dislocation quadrupole, as measured by the MP0, MPA, ACE, and MACE models. The starting points for these glides exactly match the structures in Figure \ref{fig:QuadFormation}, and the event involves the rightmost partial dislocation gliding to the right, introducing additional stacking fault between the partial cores. We see that the ACE, MACE, and MPA models give good predictions of the energy difference between the start and the end points of the migration event. We can see from the DFT singlepoint crosses at the start and end points of both panels that the ACE predicted start and endpoint are both higher in energy than the true DFT endpoint. If we account for this energy shift, the ACE, MACE, and MPA predicted barriers match very well to the DFT singlepoint predicted barrier. 

All of the models appear to underestimate the 90$^\circ$ partial dislocation quadrupole migration barrier, when compared to the DFT singlepoint crosses in the lower panel. This error can arise from both errors in the energy predictions from individual structures and also errors in the ACE predicted minimum energy path. Thus the true barrier likely lies between the estimates obtained from the DFT singlepoints and the interatomic potentials. 

\subsection{Dataset Performance}
We finally benchmark the potentials over the full InP dataset constructed in this work. We firstly calculate the binding energy, force and stress RMSEs compared with DFT over the dataset. Binding energy is the correct energy property, as the potentials have conflicting definitions of a zero energy. We calculate the binding energy for a structure by subtracting the isolated atom energy for each atom in that system from the total energy
\begin{equation}
    E_\text{Bind} = E_\text{Total} - \left( N_\text{In} E_{0;\text{In}} + N_\text{P} E_{0;\text{P}} \right)
\end{equation}
which then gives us an energy metric which is comparable across potentials. We also benchmark the inference speed of each model by finding the time taken for the model to predict energies, forces, and stresses for each structure in the dataset. The ACE, SNAP, and Vashishta models were run on a single CPU core of an Intel i9-14900K, the MP0, MPA, and MACE models were ran on a single NVIDIA RTX 4000 GPU.

\begin{table*}
\scriptsize
\caption{Comparison of Dataset Performance}
\label{tbl:Dataset}
\begin{tabular}{l||r|rrrr|rr}
\toprule
 & \begin{tabular}{r} RSCAN \\ DFT \end{tabular} & Vashishta & SNAP & MP0 & MPA & ACE & MACE \\
\midrule
\begin{tabular}{l} Binding Energy \\ (meV/Atom) \end{tabular} & - & 597.80 & 179.92 & 321.33 & 155.26 & 1.36 & 2.84 \\
\begin{tabular}{l}Forces (meV/$\text{\AA}$)\end{tabular} & - & 12,029.62 & 507.34 & 412.13 & 217.08 & 64.79 & 67.04 \\
\begin{tabular}{l}Stresses (MPa)\end{tabular} & - & 1,153.99 & 2,748.85 & 1,300.01 & 1,831.87 & 6.78 & 130.83 \\
\begin{tabular}{l} Relative Inference Time \\(mm:ss; 1 CPU or 1 GPU) \end{tabular} & - & 00:02 & 00:37 & 05:02 & 05:18 & 01:02 & 00:57 \\
\begin{tabular}{l} Million MD Steps/day \\ (1 CPU or 1 GPU; 216 atoms) \end{tabular} & - & 25.40 & 0.60 & 0.24 & 0.23 & 0.40 & 1.56 \\
\bottomrule
\end{tabular}
\end{table*}

Table \ref{tbl:Dataset} shows that the ACE and MACE models have much lower RMSEs than the other models. The MPA foundation model also performs reasonably well, as does the SNAP model, though both models have quite large stress RMSEs. From the relative inference times, we see that SNAP is around twice as fast as ACE, though both are acceptably fast. Vashishta is over an order of magnitude faster than ACE. We see that the bespoke MACE is much faster than both foundation models, due to the bespoke MACE model having a much smaller architecture.

\section{Conclusions}
\label{sec:Conc}
Through a set of benchmarks, we have shown that the ACE and MACE models developed in this work achieve a high enough accuracy to be good surrogates for DFT. Additional accuracy could likely be gained by using larger models (e.g. increasing the number of basis functions for the ACE model, or the message size in the MACE model), but the design intent was to develop models which had competitive computational cost, and could be used efficiently for large scale simulations of $>10^6$ atoms, and not just DFT-scale benchmarking.

Of the literature models tested, the MACE MPA foundation model generally shows the greatest accuracy, which bodes well for the future of the foundation model paradigm. The MPA model could be easily corrected by fine-tuning, and likely could yield accuracy comparable to the ACE and MACE models using a fine-tuning dataset much smaller than the dataset developed here, though it would be slower in inference than the MACE model due to the larger architecture. Future advances in foundation model construction, foundational datasets, and fine-tuning database design could also improve accuracy.

\section{Data Availability}
Data and scripts required to reproduce this work are provided by a GitHub repository, and corresponding Zenodo archive \cite{Zenodo}.

\section{Acknowledgments}
T.R is supported by a studentship within the Engineering and Physical Sciences Research Council supported Centre for Doctoral Training in Modelling of Heterogeneous Systems, Grant No. EP/S022848/1, with additional funding from Huawei Technologies R\&D UK.

We acknowledge usage of the ARCHER2 facility for which access was obtained via the UKCP consortium and funded by EPSRC Grants No. EP/P022561/1 and No. EP/X035891/1. Calculations were performed using the Sulis Tier 2 HPC platform hosted by the Scientific Computing Research Technology Platform at the University of Warwick. Sulis is funded by EPSRC Grant EP/T022108/1 and the HPC Midlands+ consortium. Further computing facilities were provided by the Scientific Computing Research Technology Platform of the University of Warwick.

\bibliography{ref}

@Article{SNAP,
author={Cusentino, M. A.
and Wood, M. A.
and Thompson, A. P.},
title={Explicit Multielement Extension of the Spectral Neighbor Analysis Potential for Chemically Complex Systems},
journal={The Journal of Physical Chemistry A},
year={2020},
month={Jul},
day={02},
publisher={American Chemical Society},
volume={124},
number={26},
pages={5456-5464},
issn={1089-5639},
doi={10.1021/acs.jpca.0c02450},
url={https://doi.org/10.1021/acs.jpca.0c02450}
}

@article{Vashishta,
doi = {10.1088/0953-8984/21/9/095002},
url = {https://dx.doi.org/10.1088/0953-8984/21/9/095002},
year = {2009},
month = {jan},
publisher = {},
volume = {21},
number = {9},
pages = {095002},
author = {Paulo Sergio Branicio and José Pedro Rino and Chee Kwan Gan and Hélio Tsuzuki},
title = {Interaction potential for indium phosphide: a molecular dynamics and first-principles study
of the elastic constants, generalized stacking fault and surface energies},
journal = {Journal of Physics: Condensed Matter}
}

@misc{MP0,
      title={A foundation model for atomistic materials chemistry}, 
      author={Ilyes Batatia and Philipp Benner and Yuan Chiang and Alin M. Elena and Dávid P. Kovács and Janosh Riebesell and Xavier R. Advincula and Mark Asta and Matthew Avaylon and William J. Baldwin and Fabian Berger and Noam Bernstein and Arghya Bhowmik and Samuel M. Blau and Vlad Cărare and James P. Darby and Sandip De and Flaviano Della Pia and Volker L. Deringer and Rokas Elijošius and Zakariya El-Machachi and Fabio Falcioni and Edvin Fako and Andrea C. Ferrari and Annalena Genreith-Schriever and Janine George and Rhys E. A. Goodall and Clare P. Grey and Petr Grigorev and Shuang Han and Will Handley and Hendrik H. Heenen and Kersti Hermansson and Christian Holm and Jad Jaafar and Stephan Hofmann and Konstantin S. Jakob and Hyunwook Jung and Venkat Kapil and Aaron D. Kaplan and Nima Karimitari and James R. Kermode and Namu Kroupa and Jolla Kullgren and Matthew C. Kuner and Domantas Kuryla and Guoda Liepuoniute and Johannes T. Margraf and Ioan-Bogdan Magdău and Angelos Michaelides and J. Harry Moore and Aakash A. Naik and Samuel P. Niblett and Sam Walton Norwood and Niamh O'Neill and Christoph Ortner and Kristin A. Persson and Karsten Reuter and Andrew S. Rosen and Lars L. Schaaf and Christoph Schran and Benjamin X. Shi and Eric Sivonxay and Tamás K. Stenczel and Viktor Svahn and Christopher Sutton and Thomas D. Swinburne and Jules Tilly and Cas van der Oord and Eszter Varga-Umbrich and Tejs Vegge and Martin Vondrák and Yangshuai Wang and William C. Witt and Fabian Zills and Gábor Csányi},
      year={2024},
      eprint={2401.00096},
      archivePrefix={arXiv},
      primaryClass={physics.chem-ph},
      url={https://arxiv.org/abs/2401.00096}, 
}

@article{CASTEP,
  Author    = {Clark, S. J. and Segall, M. D. and Pickard, C. J. and Hasnip, P. J.   and Probert, M. J. and Refson, K. and Payne, M.C.},
  Title     = {First principles methods using {CASTEP}},
  Journal   = {Z. Kristall.},
  Year      = {2005},
  Volume    = {220},
  Pages     = {567-570}
}

@article{RSCAN,
    author = {Bartók, Albert P. and Yates, Jonathan R.},
    title = "{Regularized SCAN functional}",
    journal = {The Journal of Chemical Physics},
    volume = {150},
    number = {16},
    pages = {161101},
    year = {2019},
    month = {04},
    issn = {0021-9606},
    doi = {10.1063/1.5094646},
    url = {https://doi.org/10.1063/1.5094646},
    eprint = {https://pubs.aip.org/aip/jcp/article-pdf/doi/10.1063/1.5094646/13514553/161101\_1\_online.pdf},
}

@article{USPP1,
  title = {Soft self-consistent pseudopotentials in a generalized eigenvalue formalism},
  author = {Vanderbilt, David},
  journal = {Phys. Rev. B},
  volume = {41},
  issue = {11},
  pages = {7892--7895},
  numpages = {0},
  year = {1990},
  month = {Apr},
  publisher = {American Physical Society},
  doi = {10.1103/PhysRevB.41.7892},
  url = {https://link.aps.org/doi/10.1103/PhysRevB.41.7892}
}

@article{USPP2,
title = {Electronic energy minimisation with ultrasoft pseudopotentials},
journal = {Computer Physics Communications},
volume = {174},
number = {1},
pages = {24-29},
year = {2006},
issn = {0010-4655},
doi = {https://doi.org/10.1016/j.cpc.2005.07.011},
url = {https://www.sciencedirect.com/science/article/pii/S0010465505004716},
author = {P.J. Hasnip and C.J. Pickard}
}

@article{NDSC1,
doi = {10.1088/2632-2153/ac9ae7},
url = {https://dx.doi.org/10.1088/2632-2153/ac9ae7},
year = {2022},
month = {dec},
publisher = {IOP Publishing},
volume = {3},
number = {4},
pages = {045031},
author = {Allen, Connor and Bartók, Albert P},
title = {Optimal data generation for machine learned interatomic potentials},
journal = {Machine Learning: Science and Technology}
}

@article{NDSC2,
  title = {Lattice dynamics and electron-phonon coupling calculations using nondiagonal supercells},
  author = {Lloyd-Williams, Jonathan H. and Monserrat, Bartomeu},
  journal = {Phys. Rev. B},
  volume = {92},
  issue = {18},
  pages = {184301},
  numpages = {9},
  year = {2015},
  month = {Nov},
  publisher = {American Physical Society},
  doi = {10.1103/PhysRevB.92.184301},
  url = {https://link.aps.org/doi/10.1103/PhysRevB.92.184301}
}

@article{matscipy, doi = {10.21105/joss.05668}, url = {https://doi.org/10.21105/joss.05668}, year = {2024}, publisher = {The Open Journal}, volume = {9}, number = {93}, pages = {5668}, author = {Petr Grigorev and Lucas Frérot and Fraser Birks and Adrien Gola and Jacek Golebiowski and Jan Grießer and Johannes L. Hörmann and Andreas Klemenz and Gianpietro Moras and Wolfram G. Nöhring and Jonas A. Oldenstaedt and Punit Patel and Thomas Reichenbach and Thomas Rocke and Lakshmi Shenoy and Michael Walter and Simon Wengert and Lei Zhang and James R. Kermode and Lars Pastewka}, title = {matscipy: materials science at the atomic scale with Python}, journal = {Journal of Open Source Software} }

@article{Ovito,
doi = {10.1088/0965-0393/18/1/015012},
url = {https://dx.doi.org/10.1088/0965-0393/18/1/015012},
year = {2009},
month = {dec},
publisher = {},
volume = {18},
number = {1},
pages = {015012},
author = {Stukowski, Alexander},
title = {Visualization and analysis of atomistic simulation data with OVITO–the Open Visualization Tool},
journal = {Modelling and Simulation in Materials Science and Engineering}
}

@article{Borcherds_1975,
doi = {10.1088/0022-3719/8/13/011},
url = {https://dx.doi.org/10.1088/0022-3719/8/13/011},
year = {1975},
month = {jul},
publisher = {},
volume = {8},
number = {13},
pages = {2022},
author = {P H Borcherds and G F Alfrey and A D B Woods and D H Saunderson},
title = {Phonon dispersion curves in indium phosphide},
journal = {Journal of Physics C: Solid State Physics}
}

@article{MaterialsProject,
    author = {Jain, Anubhav and Ong, Shyue Ping and Hautier, Geoffroy and Chen, Wei and Richards, William Davidson and Dacek, Stephen and Cholia, Shreyas and Gunter, Dan and Skinner, David and Ceder, Gerbrand and Persson, Kristin A.},
    title = {Commentary: The Materials Project: A materials genome approach to accelerating materials innovation},
    journal = {APL Materials},
    volume = {1},
    number = {1},
    pages = {011002},
    year = {2013},
    month = {07},
    issn = {2166-532X},
    doi = {10.1063/1.4812323},
    url = {https://doi.org/10.1063/1.4812323},
    eprint = {https://pubs.aip.org/aip/apm/article-pdf/doi/10.1063/1.4812323/13163869/011002\_1\_online.pdf},
}

@article{ASE,
doi = {10.1088/1361-648X/aa680e},
url = {https://dx.doi.org/10.1088/1361-648X/aa680e},
year = {2017},
month = {jun},
publisher = {IOP Publishing},
volume = {29},
number = {27},
pages = {273002},
author = {Hjorth Larsen, Ask and Jørgen Mortensen, Jens and Blomqvist, Jakob and Castelli, Ivano E and Christensen, Rune and Dułak, Marcin and Friis, Jesper and Groves, Michael N and Hammer, Bjørk and Hargus, Cory and Hermes, Eric D and Jennings, Paul C and Bjerre Jensen, Peter and Kermode, James and Kitchin, John R and Leonhard Kolsbjerg, Esben and Kubal, Joseph and Kaasbjerg, Kristen and Lysgaard, Steen and Bergmann Maronsson, Jón and Maxson, Tristan and Olsen, Thomas and Pastewka, Lars and Peterson, Andrew and Rostgaard, Carsten and Schiøtz, Jakob and Schütt, Ole and Strange, Mikkel and Thygesen, Kristian S and Vegge, Tejs and Vilhelmsen, Lasse and Walter, Michael and Zeng, Zhenhua and Jacobsen, Karsten W},
title = {The atomic simulation environment—a Python library for working with atoms},
journal = {Journal of Physics: Condensed Matter}
}

@article{MonkhorstPack,
  title = {Special points for Brillouin-zone integrations},
  author = {Monkhorst, Hendrik J. and Pack, James D.},
  journal = {Phys. Rev. B},
  volume = {13},
  issue = {12},
  pages = {5188--5192},
  numpages = {0},
  year = {1976},
  month = {Jun},
  publisher = {American Physical Society},
  doi = {10.1103/PhysRevB.13.5188},
  url = {https://link.aps.org/doi/10.1103/PhysRevB.13.5188}
}

@ARTICLE{Device1,
  author={Tomiya, S. and Hino, T. and Goto, S. and Takeya, M. and Ikeda, M.},
  journal={IEEE Journal of Selected Topics in Quantum Electronics}, 
  title={Dislocation related issues in the degradation of GaN-based laser diodes}, 
  year={2004},
  volume={10},
  number={6},
  pages={1277-1286},
  doi={10.1109/JSTQE.2004.837735}}

@article{Device2,
    author = {O’Hara, S. and Hutchinson, P. W. and Dobson, P. S.},
    title = {The origin of dislocation climb during laser operation},
    journal = {Applied Physics Letters},
    volume = {30},
    number = {8},
    pages = {368-371},
    year = {1977},
    month = {04},
    issn = {0003-6951},
    doi = {10.1063/1.89432},
    url = {https://doi.org/10.1063/1.89432},
    eprint = {https://pubs.aip.org/aip/apl/article-pdf/30/8/368/18434635/368\_1\_online.pdf},
}

@Article{DislocQuad1,
author={Ventelon, Lisa
and Willaime, F.},
title={Core structure and Peierls potential of screw dislocations in $\alpha$-Fe from first principles: cluster versus dipole approaches},
journal={Journal of Computer-Aided Materials Design},
year={2007},
month={Dec},
day={01},
volume={14},
number={1},
pages={85-94},
issn={1573-4900},
doi={10.1007/s10820-007-9064-y},
url={https://doi.org/10.1007/s10820-007-9064-y}
}

@article{DislocQuad2,
  title = {Dislocation Core Energies and Core Fields from First Principles},
  author = {Clouet, Emmanuel and Ventelon, Lisa and Willaime, F.},
  journal = {Phys. Rev. Lett.},
  volume = {102},
  issue = {5},
  pages = {055502},
  numpages = {4},
  year = {2009},
  month = {Feb},
  publisher = {American Physical Society},
  doi = {10.1103/PhysRevLett.102.055502},
  url = {https://link.aps.org/doi/10.1103/PhysRevLett.102.055502}
}

@article{DislocCyl,
doi = {10.1088/0953-8984/14/48/302},
url = {https://dx.doi.org/10.1088/0953-8984/14/48/302},
year = {2002},
month = {nov},
publisher = {},
volume = {14},
number = {48},
pages = {12673},
author = {S P Beckman and X Xu and P Specht and E R Weber and C Kisielowski and D C Chrzan},
title = {Ab initio prediction of the structure of glide set dislocation cores in GaAs},
journal = {Journal of Physics: Condensed Matter}
}

@article{QMMM1,
doi = {10.1088/0965-0393/15/3/006},
url = {https://dx.doi.org/10.1088/0965-0393/15/3/006},
year = {2007},
month = {mar},
publisher = {},
volume = {15},
number = {3},
pages = {275},
author = {Liu, Yi and Lu, Gang and Chen, Zhengzheng and Kioussis, Nicholas},
title = {An improved QM/MM approach for metals},
journal = {Modelling and Simulation in Materials Science and Engineering}
}

@article{QMMM2,
doi = {10.1088/0965-0393/19/6/065004},
url = {https://dx.doi.org/10.1088/0965-0393/19/6/065004},
year = {2011},
month = {jul},
publisher = {},
volume = {19},
number = {6},
pages = {065004},
author = {Zhao, Yi and Lu, Gang},
title = {QM/MM study of dislocation—hydrogen/helium interactions in \textalpha-Fe},
journal = {Modelling and Simulation in Materials Science and Engineering}
}

@article{QMMM3,
  title = {Hybrid quantum/classical study of hydrogen-decorated screw dislocations in tungsten: Ultrafast pipe diffusion, core reconstruction, and effects on glide mechanism},
  author = {Grigorev, Petr and Swinburne, Thomas D. and Kermode, James R.},
  journal = {Phys. Rev. Mater.},
  volume = {4},
  issue = {2},
  pages = {023601},
  numpages = {10},
  year = {2020},
  month = {Feb},
  publisher = {American Physical Society},
  doi = {10.1103/PhysRevMaterials.4.023601},
  url = {https://link.aps.org/doi/10.1103/PhysRevMaterials.4.023601}
}

@article{DislocDipole,
author = {Søren L. Frederiksen and Karsten W. Jacobsen},
title = {Density functional theory studies of screw dislocation core structures in bcc metals},
journal = {Philosophical Magazine},
volume = {83},
number = {3},
pages = {365--375},
year = {2003},
publisher = {Taylor \& Francis},
doi = {10.1080/0141861021000034568},
URL = {https://doi.org/10.1080/0141861021000034568},
eprint = {https://doi.org/10.1080/0141861021000034568}
}

@article{DislocGreens1,
author = {S. Rao and C. Hernandez and J. P. Simmons and T. A. Parthasarathy and C. Woodward},
title = {Green's function boundary conditions in two-dimensional and three-dimensional atomistic simulations of dislocations},
journal = {Philosophical Magazine A},
volume = {77},
number = {1},
pages = {231--256},
year = {1998},
publisher = {Taylor \& Francis},
doi = {10.1080/01418619808214240},
URL = {https://doi.org/10.1080/01418619808214240},
eprint = {https://doi.org/10.1080/01418619808214240}
}

@article{DislocGreens2,
  title = {Lattice Green function for extended defect calculations: Computation and error estimation with long-range forces},
  author = {Trinkle, D. R.},
  journal = {Phys. Rev. B},
  volume = {78},
  issue = {1},
  pages = {014110},
  numpages = {11},
  year = {2008},
  month = {Jul},
  publisher = {American Physical Society},
  doi = {10.1103/PhysRevB.78.014110},
  url = {https://link.aps.org/doi/10.1103/PhysRevB.78.014110}
}

@article{Alexandria,
title = {Improving machine-learning models in materials science through large datasets},
journal = {Materials Today Physics},
volume = {48},
pages = {101560},
year = {2024},
issn = {2542-5293},
doi = {https://doi.org/10.1016/j.mtphys.2024.101560},
url = {https://www.sciencedirect.com/science/article/pii/S2542529324002360},
author = {Jonathan Schmidt and Tiago F.T. Cerqueira and Aldo H. Romero and Antoine Loew and Fabian Jäger and Hai-Chen Wang and Silvana Botti and Miguel A.L. Marques}
}

@misc{sAlex,
      title={Open Materials 2024 (OMat24) Inorganic Materials Dataset and Models}, 
      author={Luis Barroso-Luque and Muhammed Shuaibi and Xiang Fu and Brandon M. Wood and Misko Dzamba and Meng Gao and Ammar Rizvi and C. Lawrence Zitnick and Zachary W. Ulissi},
      year={2024},
      eprint={2410.12771},
      archivePrefix={arXiv},
      primaryClass={cond-mat.mtrl-sci},
      url={https://arxiv.org/abs/2410.12771}, 
}

@article{DislocPD1,
  title = {Theory of Dislocation Mobility in Semiconductors},
  author = {Celli, V. and Kabler, M. and Ninomiya, T. and Thomson, R.},
  journal = {Phys. Rev.},
  volume = {131},
  issue = {1},
  pages = {58--72},
  numpages = {0},
  year = {1963},
  month = {Jul},
  publisher = {American Physical Society},
  doi = {10.1103/PhysRev.131.58},
  url = {https://link.aps.org/doi/10.1103/PhysRev.131.58}
}

@article{DislocPD2,
doi = {10.1088/0034-4885/33/1/303},
url = {https://dx.doi.org/10.1088/0034-4885/33/1/303},
year = {1970},
month = {jan},
publisher = {},
volume = {33},
number = {1},
pages = {101},
author = {R Bullough and R C Newman},
title = {The kinetics of migration of point defects to dislocations},
journal = {Reports on Progress in Physics}
}

@article{DislocPD3,
author = {R. Vardya and S. Mahajan},
title = {Mechanism of dislocation climb in binary and mixed III-V semiconductors},
journal = {Philosophical Magazine A},
volume = {71},
number = {3},
pages = {465--472},
year = {1995},
publisher = {Taylor \& Francis},
doi = {10.1080/01418619508244462},
URL = {https://doi.org/10.1080/01418619508244462},
eprint = {https://doi.org/10.1080/01418619508244462}
}

@article{DiamondPartials,
  title = {Dislocations in diamond: Core structures and energies},
  author = {Blumenau, A. T. and Heggie, M. I. and Fall, C. J. and Jones, R. and Frauenheim, T.},
  journal = {Phys. Rev. B},
  volume = {65},
  issue = {20},
  pages = {205205},
  numpages = {8},
  year = {2002},
  month = {May},
  publisher = {American Physical Society},
  doi = {10.1103/PhysRevB.65.205205},
  url = {https://link.aps.org/doi/10.1103/PhysRevB.65.205205}
}

@article{InPPartials,
title = {The characterization of misfit dislocations at {100} heterojunctions in III–V compound semiconductors},
journal = {Acta Metallurgica},
volume = {37},
number = {10},
pages = {2779-2793},
year = {1989},
issn = {0001-6160},
doi = {https://doi.org/10.1016/0001-6160(89)90312-X},
url = {https://www.sciencedirect.com/science/article/pii/000161608990312X},
author = {B.C. {De Cooman} and C.B. Carter and Chan {Kam Toi} and J.R. Shealy}
}

@article{ACEpotentials,
    author = {Witt, William C. and van der Oord, Cas and Gelžinytė, Elena and Järvinen, Teemu and Ross, Andres and Darby, James P. and Ho, Cheuk Hin and Baldwin, William J. and Sachs, Matthias and Kermode, James and Bernstein, Noam and Csányi, Gábor and Ortner, Christoph},
    title = {ACEpotentials.jl: A Julia implementation of the atomic cluster expansion},
    journal = {The Journal of Chemical Physics},
    volume = {159},
    number = {16},
    pages = {164101},
    year = {2023},
    month = {10},
    issn = {0021-9606},
    doi = {10.1063/5.0158783},
    url = {https://doi.org/10.1063/5.0158783},
}

@article{ACE,
  title = {Atomic cluster expansion for accurate and transferable interatomic potentials},
  author = {Drautz, Ralf},
  journal = {Phys. Rev. B},
  volume = {99},
  issue = {1},
  pages = {014104},
  numpages = {15},
  year = {2019},
  month = {Jan},
  publisher = {American Physical Society},
  doi = {10.1103/PhysRevB.99.014104},
  url = {https://link.aps.org/doi/10.1103/PhysRevB.99.014104}
}

@inproceedings{MACE,
 author = {Batatia, Ilyes and Kovacs, David P and Simm, Gregor and Ortner, Christoph and Csanyi, Gabor},
 booktitle = {Advances in Neural Information Processing Systems},
 editor = {S. Koyejo and S. Mohamed and A. Agarwal and D. Belgrave and K. Cho and A. Oh},
 pages = {11423--11436},
 publisher = {Curran Associates, Inc.},
 title = {MACE: Higher Order Equivariant Message Passing Neural Networks for Fast and Accurate Force Fields},
 url = {https://proceedings.neurips.cc/paper_files/paper/2022/file/4a36c3c51af11ed9f34615b81edb5bbc-Paper-Conference.pdf},
 volume = {35},
 year = {2022}
}

@inproceedings{SWA,
  author       = {Pavel Izmailov and
                  Dmitrii Podoprikhin and
                  Timur Garipov and
                  Dmitry P. Vetrov and
                  Andrew Gordon Wilson},
  editor       = {Amir Globerson and
                  Ricardo Silva},
  title        = {Averaging Weights Leads to Wider Optima and Better Generalization},
  booktitle    = {Proceedings of the Thirty-Fourth Conference on Uncertainty in Artificial
                  Intelligence, {UAI} 2018, Monterey, California, USA, August 6-10,
                  2018},
  pages        = {876--885},
  publisher    = {{AUAI} Press},
  year         = {2018},
  url          = {http://auai.org/uai2018/proceedings/papers/313.pdf},
  bibsource    = {dblp computer science bibliography, https://dblp.org}
}

@dataset{Zenodo,
  author = {Rocke, Thomas and Kermode, James},
  title = {InP MLIP Testing Framework},
  month = {8},
  year = {2025},
  publisher = {Zenodo},
  doi = {10.5281/zenodo.16904066},
  url = {https://doi.org/10.5281/zenodo.16904066},
}

@article{PDForm2,
  title = {First-principles calculations for point defects in solids},
  author = {Freysoldt, Christoph and Grabowski, Blazej and Hickel, Tilmann and Neugebauer, J\"org and Kresse, Georg and Janotti, Anderson and Van de Walle, Chris G.},
  journal = {Rev. Mod. Phys.},
  volume = {86},
  issue = {1},
  pages = {253--305},
  numpages = {53},
  year = {2014},
  month = {Mar},
  publisher = {American Physical Society},
  doi = {10.1103/RevModPhys.86.253},
  url = {https://link.aps.org/doi/10.1103/RevModPhys.86.253}
}

@article{Vurgaftman,
author = {Vurgaftman, I and Meyer, Jerry and Ram-Mohan, Ramdas},
year = {2001},
month = {06},
pages = {5815-5875},
title = {Band parameters for III-V compound semiconductors and their alloys},
volume = {89},
journal = {J. Appl. Phys.},
doi = {10.1063/1.1368156}
}

@article{Cunnel,
author = {Cunnel, F. A. and Saker, E. W. },
year = {1957},
title = {Properties of III-V compound semiconductors},
volume = {2},
journal = {Progress in Solids},
pages = {37-65}
}

@Article{Vasilev,
author={Vasil'ev, V. P.
and Gachon, J.-C.},
title={Thermodynamic properties of InP},
journal={Inorganic Materials},
year={2006},
month={Nov},
day={01},
volume={42},
number={11},
pages={1171-1175},
issn={1608-3172},
doi={10.1134/S002016850611001X},
url={https://doi.org/10.1134/S002016850611001X}
}

@article{Nichols,
title = {Elastic anharmonicity of InP: Its relationship to the high pressure transition},
journal = {Solid State Communications},
volume = {36},
number = {8},
pages = {667-669},
year = {1980},
issn = {0038-1098},
doi = {https://doi.org/10.1016/0038-1098(80)90205-7},
url = {https://www.sciencedirect.com/science/article/pii/0038109880902057},
author = {D.N. Nichols and D.S. Rimai and R.J. Sladek}
}

@article{Hickernell,
    author = {Hickernell, F. S. and Gayton, W. R.},
    title = {Elastic Constants of Single‐Crystal Indium Phosphide},
    journal = {Journal of Applied Physics},
    volume = {37},
    number = {1},
    pages = {462-462},
    year = {1966},
    month = {01},
    issn = {0021-8979},
    doi = {10.1063/1.1707886},
    url = {https://doi.org/10.1063/1.1707886},
}
\end{document}